%% file: ms.tex
\documentclass[twocolumn,tighten]{aastex701}
\hypersetup{linkcolor=blue,citecolor=blue,filecolor=blue,urlcolor=blue}
\usepackage{amsmath}
\usepackage{natbib}
\usepackage{enumitem}

\newcommand{\hdone}{\object[HD 196944]{HD~196944}}
\newcommand{\hdtwo}{\object[HD 222925]{HD~222925}}

\newcommand{\loggf}{\mbox{$\log(gf)$}}
\newcommand{\kmsec}{\mbox{km~s$^{\rm -1}$}}
\newcommand{\logg}{\mbox{log~{\it g}}}
\newcommand{\msun}{\mbox{$M_{\odot}$}}
\newcommand{\teff}{\mbox{$T_{\rm eff}$}}
\newcommand{\vt}{\mbox{$v_{\rm t}$}}
\newcommand{\rpro}{\mbox{{\it r}-process}}
\newcommand{\spro}{\mbox{{\it s}-process}}
\newcommand{\ipro}{\mbox{{\it i}-process}}

\newcommand{\logeps}[1]{$\log\varepsilon$(#1)}

\newcommand{\cemps}{\mbox{CEMP-\textit{s}}}
\newcommand{\cemprs}{\mbox{CEMP-\textit{r/s}}}

\shorttitle{New UV Abundances in HD 196944}
\shortauthors{Roederer et al.}
\accepted{for publication in the Astrophysical Journal}

\begin{document}

\title{%
Abundances of Rarely Detected \textit{s}-process Elements
Derived from the Ultraviolet Spectrum \\
of the \textit{s}-process-enhanced Metal-poor Star HD~196944\footnote{%
Based on observations made with the NASA/ESA 
Hubble Space Telescope, 
obtained at the Space Telescope Science Institute (STScI), which is 
operated by the Association of Universities for 
Research in Astronomy, Inc.\ (AURA) under NASA contract NAS~5-26555.
These observations are associated with programs GO-12554 and GO-14765.
This paper also includes data gathered at 
the McDonald Observatory of The University of Texas at Austin, and
the 6.5~meter Magellan Telescopes located at Las Campanas Observatory, Chile.
}
}

\author[0000-0001-5107-8930]{Ian U.\ Roederer}
\affiliation{%
Department of Physics, North Carolina State University,
Raleigh, NC 27695, USA}
\email{iuroederer@ncsu.edu}

\author[0000-0003-4479-1265]{Vinicius M.\ Placco}
\affiliation{%
NSF NOIRLab, 
Tucson, AZ 85719, USA}
\email{vinicius.placco@noirlab.edu}

\author[0000-0002-3625-6951]{Amanda I.\ Karakas}
\affiliation{%
Monash Centre for Astrophysics, School of Physics and Astronomy, 
Monash University, VIC 3800, Australia}
\email{Amanda.Karakas@monash.edu}

\author[0000-0001-8582-0910]{Elizabeth A.\ Den Hartog}
\affiliation{%
Department of Physics, University of Wisconsin-Madison,
Madison, WI 53706, USA}
\email{eadenhar@facstaff.wisc.edu}

\author[0000-0003-4573-6233]{Timothy C.\ Beers}
\affiliation{%
Department of Physics, University of Notre Dame, 
Notre Dame, IN 46556, USA}
\email{tbeers@nd.edu}

\begin{abstract}

We present an analysis of the
heavy-element abundances
of \mbox{HD~196944},
a carbon-enhanced metal-poor (CEMP) star 
enriched with elements produced by the
slow neutron-capture process ({\it s}-process).
We obtained a new high-resolution ultraviolet (UV)
spectrum of this star, the UV-brightest known \mbox{CEMP-{\it s}} star,
with the Space Telescope Imaging Spectrograph
on board the Hubble Space Telescope.
This spectrum extends deeper into the UV
(2029 $\leq \lambda \leq$ 2303~\AA)
than previous studies of any \mbox{CEMP-{\it s}} star.
When combined with previous UV and optical analysis,
a total of 35 elements heavier than zinc ($Z$ = 30)
can be detected in \mbox{HD~196944},
and upper limits are available for
nine other heavy elements.
The abundances can be well fit
by models of {\it s}-process nucleosynthesis operating in a 
low-mass companion star 
that evolved through the asymptotic giant branch phase
and transferred heavy elements to \mbox{HD~196944}.
This {\it s}-process event did not contribute substantially
to the Ga, Ge, or As abundances (31 $\leq Z \leq$ 33).
Our results demonstrate that UV spectroscopy 
can greatly expand the inventory
of heavy elements detectable in \mbox{CEMP-{\it s}} stars.

\end{abstract}



\section{Introduction}
\label{intro}

\setcounter{footnote}{12}

Understanding the cosmic origins of elements 
listed all across the periodic table
remains an active area of study.
Elements heavier than the iron group, 
those with atomic numbers ($Z$) $>$~30,
are primarily produced by neutron-capture reactions.
Rapid-process (\rpro) reactions occur when the 
neutron captures occur faster than the rate of intervening $\beta$ decays, 
and 
slow-process (\spro) reactions occur when the 
neutron captures are slower than the $\beta$ decays
(e.g., \citealt{burbidge57}).

The elemental-abundance patterns observed in stars reflect these
different scenarios.
For example, 
overabundances, or peaks, in the \rpro\ abundance
distribution are found at mass numbers ($A$) of $\approx$~80, 130, and 195,
whereas the three peaks in the \spro\ abundance distribution 
are found at $A \approx$~90, 138, and 208.
These peaks occur in response to the small neutron-capture cross sections
at filled nuclear shells with
50, 82, and 126 neutrons.
The \rpro\ encounters these peaks far from stability
at neutron-rich nuclei that subsequently $\beta^{-}$ decay
to stability, while
the \spro\ encounters them at stable isotopes.
These differences result in the distinct 
abundance patterns of each of these processes.

Most late-type (FGK) stars acquired their
neutron-capture elements at birth or by accretion from a companion star.
It is generally easy to determine
whether some form of \rpro\ or some form of \spro\
dominated the nucleosynthesis of elements
in a particular star.
This information is encoded in
the abundance ratios of relatively small numbers of key elements,
such as Sr, Ba, and Eu
(e.g., \citealt{beers05}).
When abundances of additional elements are considered,
the rich diversity of variations on the simple \textit{r}- and \spro\
categories of nucleosynthesis comes into view
(see reviews by, e.g., \citealt{cowan21} and \citealt{lugaro23}).

Each spectral domain offers access to absorption lines
of different elements.
The blue region of the optical spectrum
and the ultraviolet (UV) region are especially rich in
absorption lines of neutron-capture elements.
The vast majority of studies of neutron-capture elements
in stars have used optical spectroscopy
(e.g., \citealt{burris00,battistini16,sakari18north}),
although near-infrared spectroscopy has also
made major contributions in recent years
(e.g., \citealt{matsunaga20,salessilva22,nandakumar24}).
The samples of stars accessible through 
UV spectroscopy are still many orders
of magnitude smaller
(e.g., \citealt{cowan05,roederer12d,peterson20}), 
because the high-resolution UV spectrographs on the 
Hubble Space Telescope (HST)
remain the only source of observational material.

Most previous studies of neutron-capture elements in the 
UV spectra of metal-poor stars have examined 
\rpro-enhanced stars
(e.g., \citealt{sneden03a,siqueiramello13,shah24}),
stars with relatively low enhancements of neutron-capture elements
(e.g., \citealt{sneden98,roederer10b,roederer16d,placco14bdp44}),
or stars with more complex interpretations
(e.g., \citealt{peterson11,roederer12c,roederer16c}).
Only one study \citep{placco15cemps}
has focused on carbon-enhanced metal-poor (CEMP) stars
with \spro\ enhancements (\cemps; \citealt{beers05}).
This bias in the literature reflects the
scientific interests of the investigators who have collected and
analyzed UV spectra over the last 30~years,
not any inherent difficulty
in obtaining UV spectra for \spro-enhanced stars.

Here, we present an analysis of new UV spectroscopic observations
of one of the \cemps\ stars examined by \citet{placco15cemps},
the \textit{G}-type red giant \hdone.
The first analysis of its detailed chemical composition 
based on high-resolution optical spectra 
was published by \citet{zacs98}, 
who confirmed previous suggestions of enhanced 
carbon and oxygen abundances
([C/Fe] and [O/Fe] both $> +1$)
relative to its low metallicity ([Fe/H] = $-2.4$).
That study also confirmed enhanced levels of \spro\ elements
at both the first and second \spro\ peaks
([Y/Fe] = $+$0.58 and [Ba/Fe] = $+$1.56).
\citet{vaneck01} discovered the enormous overabundance of the 
third \spro-peak element lead in \hdone\ ([Pb/Fe] = $+$2.15),
which had been anticipated by \citet{gallino98}.
\hdone\ was the subject of many more optical spectroscopic studies
in subsequent years, including ones by
\citet{aoki02pb,aoki07cemp}, \citet{masseron10},
\citet{roederer14c}, \citet{karinkuzhi21}, and \citet{contursi24}.
These studies have shown that the nitrogen and
sodium abundances are both enhanced relative to
other metal-poor stars
([N/Fe] $\approx +$1.3, [Na/Fe] $\approx +$0.8).
Otherwise, its light-element abundance pattern ($Z \leq$ 30) 
is typical for its metallicity.
These studies have derived abundances for $\approx$~15
elements heavier than Zn in \hdone.

The enhancements observed in \hdone\
are commonly understood to be the result of
\spro\ nucleosynthesis
in a more-massive companion star
that passed through the asymptotic giant branch (AGB) phase of evolution.
This star transferred mass, including the freshly produced C, N, O, Na, and
\spro\ elements, to \hdone.
Radial velocity measurements
reveal that \hdone\ is the more luminous star in a binary system
\citep{lucatello05}.
\citet{placco15cemps} calculated the first orbital solution
for this system, finding a period of 1325 $\pm$~9~d,
and \citet{karinkuzhi21} quoted an updated orbital period of 1294~d.

The abundance pattern of \hdone\
can be compared with predictions for
\spro\ nucleosynthesis in AGB stars
of various masses, metallicities, neutron exposures,
and dilution factors.
\citet{bisterzo11} found the AGB had 
an initial mass of 1.5~\msun,
the same metallicity as \hdone,
a neutron exposure less than the standard prescription, and
a dilution factor of 1\%.  
\citet{abate15b} fit a model from their grid of binary evolution models:\
initial mass of the (primary) companion star of 1.40$^{+0.25}_{-0.45}$~\msun,
initial mass of \hdone\ of 0.81$^{0.07}_{-0.06}$~\msun, 
and initial orbital period of 5840$^{+4080}_{-4460}$~d.
\citet{placco15cemps}\ applied their orbital period constraint
and the abundance pattern
to select the best-fitting model from the \citeauthor{abate15b} grid:\
initial mass of the companion star of 0.9~\msun,
initial mass of \hdone\ of 0.86~\msun, 
and initial orbital period of 1640~d.
The binary nature of \hdone\ is settled,
but the initial characteristics 
of the now-evolved putative white dwarf companion
remain unsettled.

The goal of the present study is to 
derive abundances of additional \spro\ elements in \hdone\
by pushing deeper into the UV than was done previously,
to wavelengths as short as $\approx$~2000~\AA.
We collect new UV spectra of \hdone\ with HST
(Section~\ref{obs}).
We use these data to detect elements 
that have not been detected previously in \cemps\ stars
(Section~\ref{analysis}).
A total of 35~elements heavier than Zn ($Z >$~30)
are detectable in the UV and optical spectra of \hdone\
(Section~\ref{results}).
We discuss these results in the context of 
models of \spro\ nucleosynthesis operating 
in low-metallicity AGB stars and related matters
(Section~\ref{discussion}), and
we summarize our conclusions (Section~\ref{conclusions}).

\section{Observations}
\label{obs}

\subsection{New UV Spectra}

\hdone\ was observed using the Space Telescope Imaging Spectrograph
(STIS; \citealt{kimble98,woodgate98}) on HST.~
These observations were made using the E230H
echelle grating centered at $\lambda$2163, 
the 0\farcs2~$\times$~0\farcs09 slit, and
the near-UV Multianode Microchannel Array (MAMA) detector.
This setup produces spectra with a resolving power of
$R \equiv \lambda/\Delta\lambda =$~114,000
and wavelength coverage from 2029--2303~\AA.~
The observations were made over the course of 40 orbits 
spread across 14~visits
from 2017 September 02 to 2017 November 16.
Integration times of individual exposures
ranged from 1806~s to 2961~s, for a total
integration time of 102,270~s (28.4~hr).
The spectra were processed automatically by the 
CALSTIS software package and downloaded from the
Mikulski Archive for Space Telescopes (MAST).
We shifted all individual observations to a common rest velocity
before co-adding.
We co-added and normalized the spectrum 
using tools in the 
Image Reduction and Analysis Facility 
(IRAF; \citealt{tody86,tody93,fitzpatrick24}) ``onedspec'' package.
The signal-to-noise (S/N) ratios per pixel 
in the co-added spectrum
are approximately
10/1 at 2030~\AA, 
25/1 at 2100~\AA,
40/1 at 2200~\AA,
and
50/1 at 2300~\AA.%
\footnote{These data can be obtained from the MAST archive at this link:\
\dataset[10.17909/ydpv-ma49]{https://dx.doi.org/10.17909/ydpv-ma49}}

\subsection{Previously Published Spectra}
\label{sec:previousspec}

We also reanalyze a few lines in
high-resolution optical spectra of \hdone\
collected and published previously.
One high-resolution ($R$ = 120,000) optical spectrum
was obtained in 2006 using the 2dCoud\'{e} echelle spectrograph
\citep{tull95} on the 2.7~m Harlan J.\ Smith Telescope
at McDonald Observatory.
This spectrum covers 4200 $\leq \lambda \leq$ 6640~\AA, 
with gaps in coverage, at high S/N (160/1 at $\lambda$ = 4200~\AA).~
This spectrum was previously analyzed and published by
\citet{roederer08a}.

We also reanalyzed a 
second high-resolution ($R$ = 41,000 for $\lambda < 5000$~\AA; 
$R$ = 35,000 for $\lambda > 5000$~\AA)
high-S/N (230/1 at $\lambda$ = 4550~\AA)
optical (3330 $\leq \lambda \leq$ 8000~\AA)
spectrum of \hdone.
This spectrum was collected in 2009 using the 
Magellan Inamori Kyocera Echelle (MIKE) spectrograph \citep{bernstein03}
at the Landon Clay Telescope (Magellan~II) at Las Campanas Observatory.
This spectrum was previously analyzed and published by 
\citet{roederer14c}.

We also reanalyzed portions of the STIS/E230M spectrum of \hdone.
This spectrum covers 2280 $\leq \lambda \leq$ 3070~\AA\
at $R$ = 30,000 and S/N = 45/1 at $\lambda$ = 2300~\AA.~
This UV spectrum was previously analyzed and published by 
\citet{placco15cemps}.

\section{Analysis}
\label{analysis}

We adopt the standard nomenclature 
for elemental abundances and ratios.
The abundance of an element X is defined
as the number of X atoms per 10$^{12}$ H atoms,
$\log\varepsilon$(X)~$\equiv \log_{10}(N_{\rm X}/N_{\rm H})+$12.0.
The abundance ratio of elements X and Y relative to the
Solar ratio is defined as
[X/Y] $\equiv \log_{10} (N_{\rm X}/N_{\rm Y}) - \log_{10} (N_{\rm X}/N_{\rm Y})_{\odot}$.
We adopt the Solar photospheric abundances of \citet{lodders09}.
By convention,
abundances or ratios denoted with the ionization state
are understood to be
the total elemental abundance, as derived from transitions of
that particular ionization state 
after Saha ionization corrections have been applied.

\subsection{Stellar Parameters}
\label{sec:params}

We adopt the model atmosphere for \hdone\ derived
by \citet{placco15cemps}
so that we can add our newly derived abundances to their scale
without needing to rederive abundances from the 321~lines analyzed 
in that study.
That model atmosphere has 
effective temperature (\teff) $=$ 5170~$\pm$~100~K,
log of the surface gravity (\logg) $=$ 1.60~$\pm$~0.25 [cgs],
microturbulent velocity (\vt) $=$ 1.55~$\pm$~0.10~\kmsec,
and
model metallicity ([M/H]) $= -$2.41~$\pm$~0.25.
These values identify \hdone\ as a core-He-burning star.
This model is interpolated from a one-dimensional,
hydrostatic model atmosphere from the $\alpha$-enhanced
ATLAS9 grid of models \citep{castelli04}.
We adopt a macroturbulent broadening of 6.8~\kmsec\
\citep{roederer08a}
when generating synthetic spectra.

We derive abundances 
using a recent version of the 
line analysis software MOOG
(\citealt{sneden73}; 2017 version).
MOOG assumes that local thermodynamic equilibrium (LTE)
holds in the line-forming layers of the atmosphere.
This version of the code treats 
Rayleigh scattering, which affects 
the continuous opacity at shorter wavelengths,
as isotropic, coherent scattering,
as described in \citet{sobeck11}.
We adopt 
damping constants for collisional broadening
with neutral H from \citet{barklem00h}
and \citet{barklem05feii}, when available,
otherwise
we adopt the standard \citet{unsold55} recipe.

\subsection{Iron}
\label{felinelist}

We use spectrum synthesis matching to 
check the iron abundance derived
from UV Fe~\textsc{i} and Fe~\textsc{ii} lines.
We generate line lists using the 
LINEMAKE software \citep{placco21linemake}.
LINEMAKE creates an initial line list based 
on the \citet{kurucz11} lists.
Next, LINEMAKE
updates the atomic data in these lists
using results for
wavelengths, \loggf\ values, and 
any hyperfine splitting (HFS) structure or isotope shifts (IS).~
Sources of these data include the Fe-group species summarized
in \citet{lawler17review}, 
subsequent analyses (e.g., \citealt{denhartog23}), 
and other strong transitions 
in the 
National Institute of Standards and Technology (NIST)
Atomic Spectra Database (ASD; \citealt{kramida21} version
for the analysis presented here).
We further update these lists with the Fe~\textsc{i}
transitions recently identified by \citet{peterson15,peterson22} and
\citet{peterson17}.

We consider all 47 Fe~\textsc{i} lines in this wavelength range
with laboratory \loggf\ values presented by
\citet{belmonte17} 
or with grades D$+$ or better in the 
NIST ASD.~
We find 11 of these lines to be relatively
unblended and not too saturated,
and they yield an unweighted (weighted) [Fe/H]~$= -2.38 \pm 0.07$
($-2.41 \pm 0.06$) with standard deviation ($\sigma$) of 0.22~dex.
These lines are all weak transitions that connect to 
low-lying levels 
(excitation potential, E.P., $<$~0.2~eV; Table~\ref{linetab}).
Similarly, we consider all 263 Fe~\textsc{ii} lines
in this wavelength range with laboratory \loggf\ values
presented by \citet{denhartog19}
or with grades D or better in the NIST ASD.~
We find 67 of these lines to be relatively
unblended and not too saturated,
and they yield an unweighted (weighted) [Fe/H]~$= -2.49 \pm 0.03$
($-2.53 \pm 0.03$) with $\sigma = 0.23$~dex.
These abundances agree 
with the [Fe/H] abundance derived
from optical lines by \citet{placco15cemps},
[Fe/H]~$= -2.41 \pm 0.18$.
This agreement indicates that 
our method of deriving abundances
from lines at these UV wavelengths 
matches the abundance scale established by the optical lines.

\input{tab1-stub}

\subsection{Other Elements}

\begin{figure*}
\begin{center}
\includegraphics[angle=0,width=3.35in]{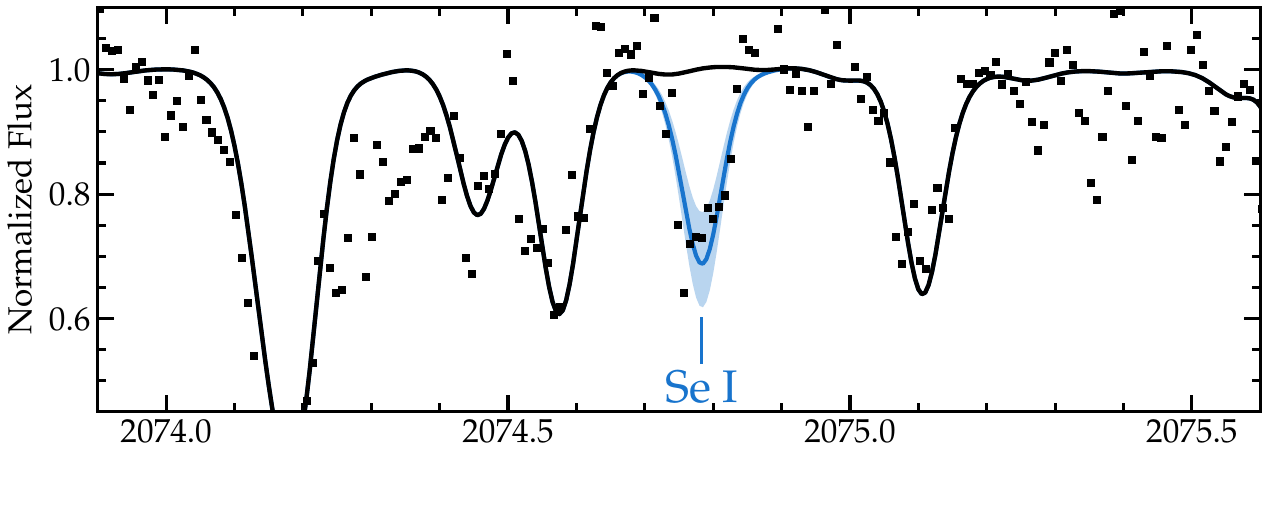}
\hspace*{0.1in}
\includegraphics[angle=0,width=3.35in]{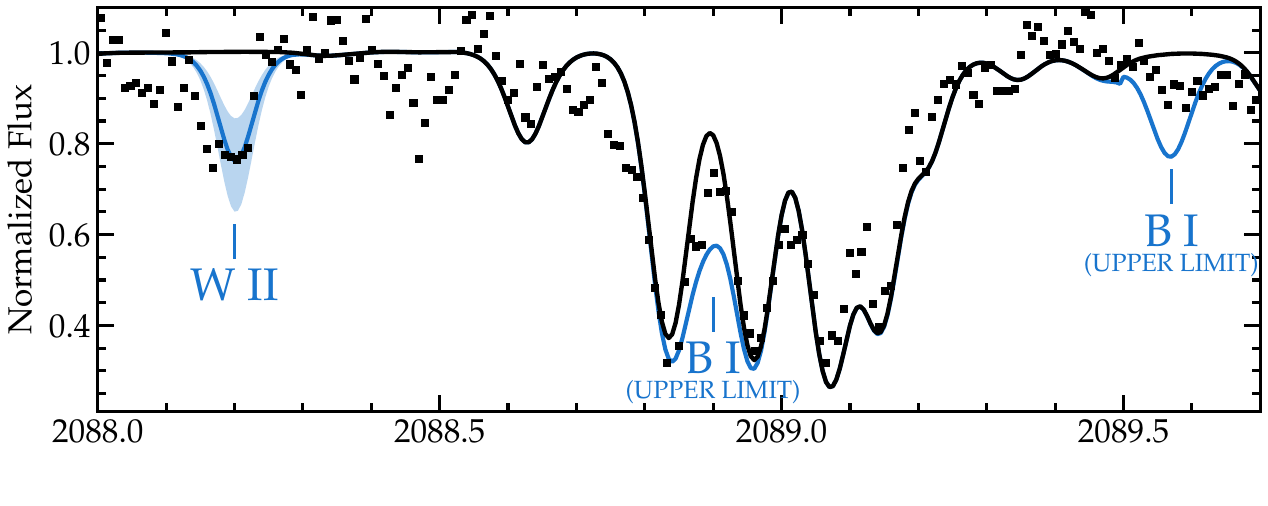} \\
\vspace*{-0.1in}
\includegraphics[angle=0,width=3.35in]{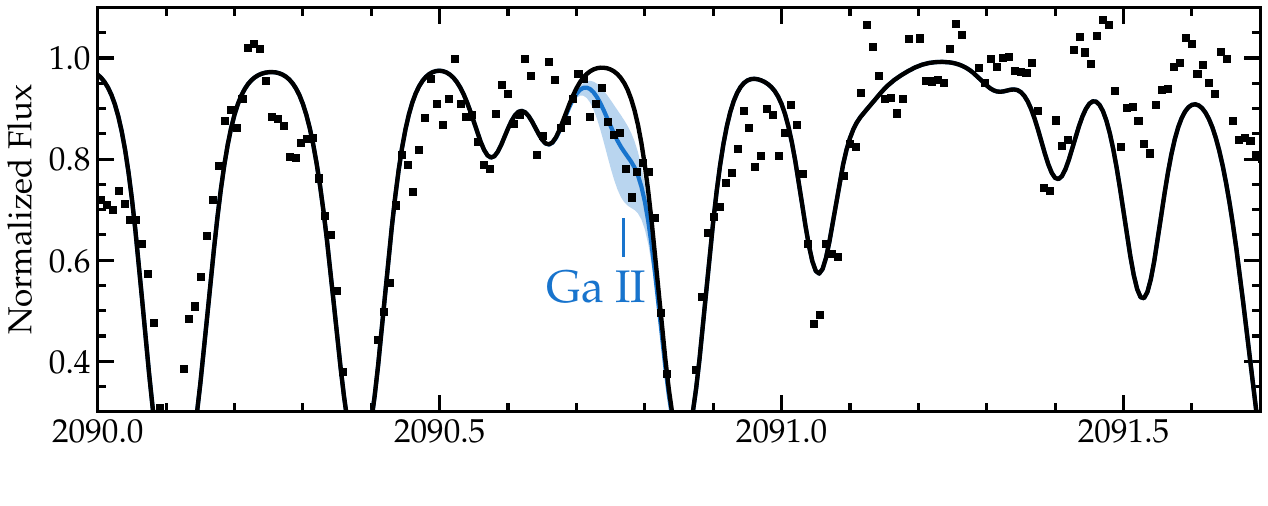}
\hspace*{0.1in}
\includegraphics[angle=0,width=3.35in]{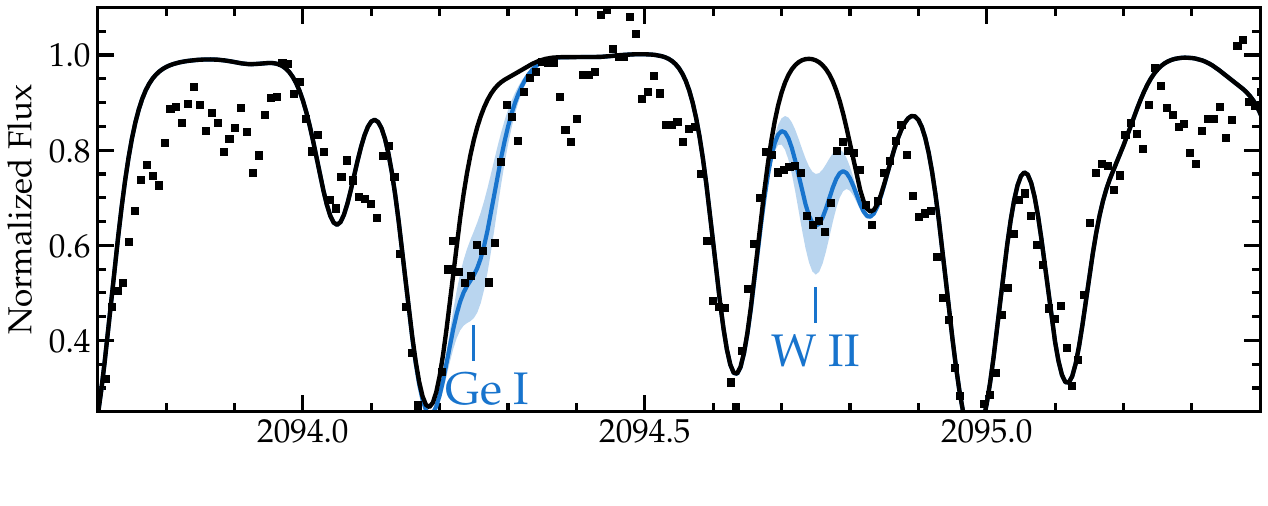} \\
\vspace*{-0.1in}
\includegraphics[angle=0,width=3.35in]{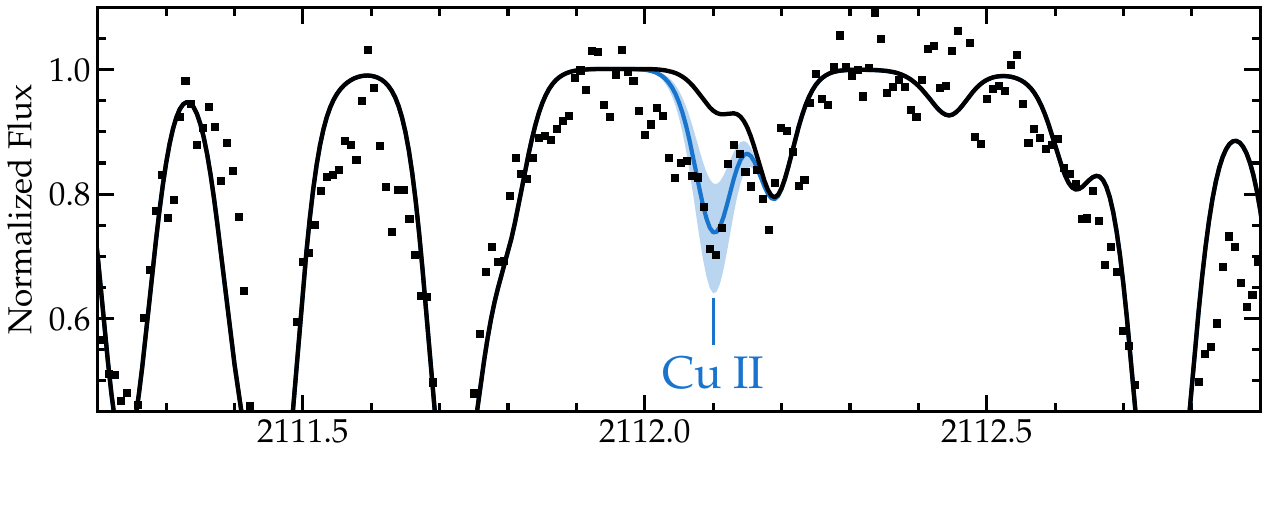}
\hspace*{0.1in}
\includegraphics[angle=0,width=3.35in]{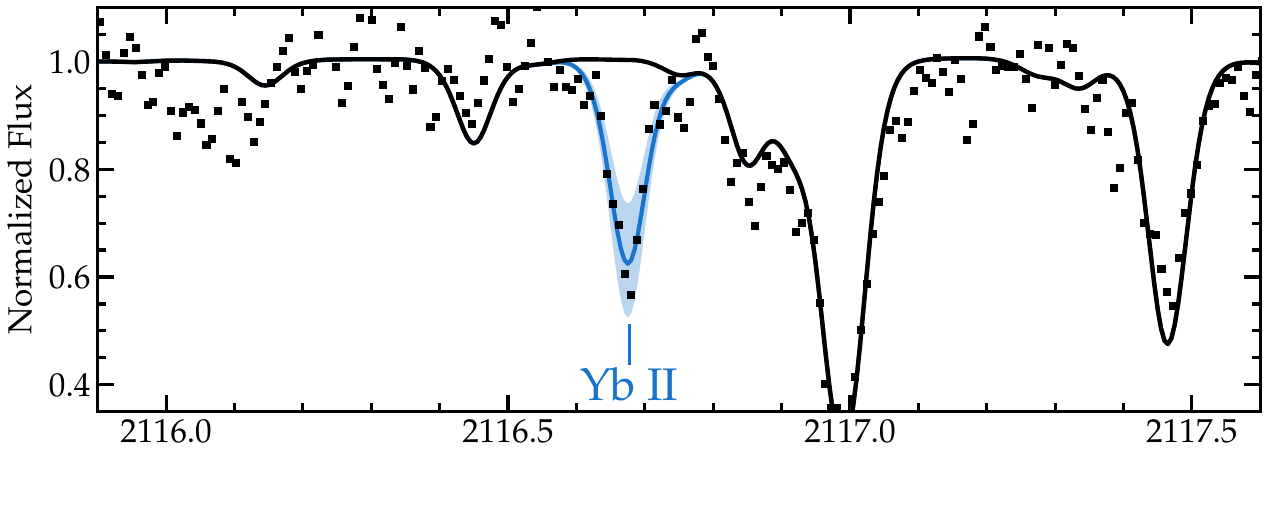} \\
\vspace*{-0.1in}
\includegraphics[angle=0,width=3.35in]{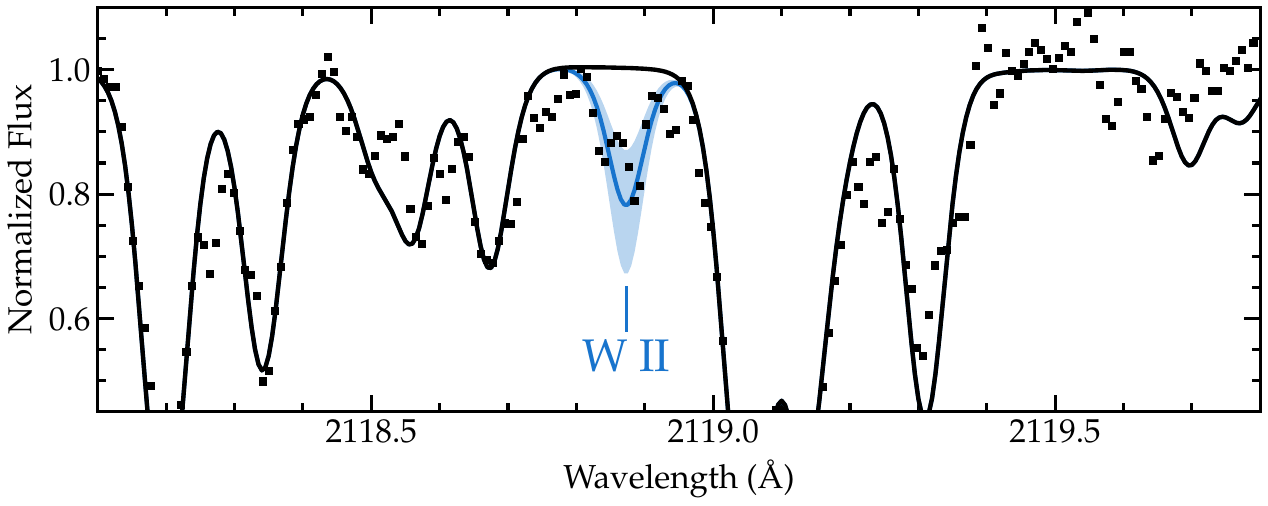}
\hspace*{0.1in}
\includegraphics[angle=0,width=3.35in]{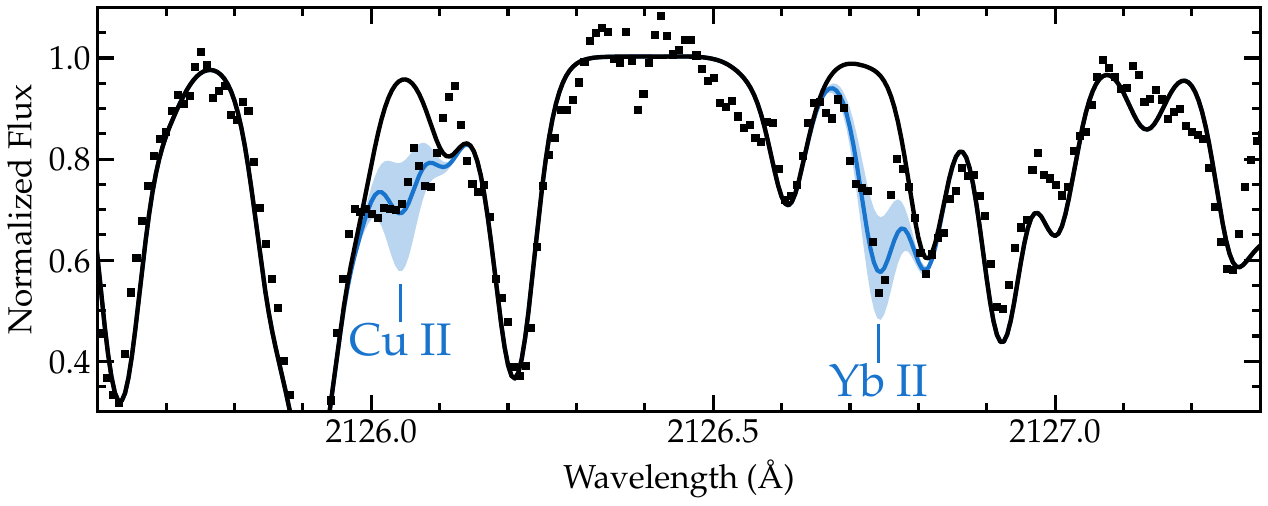} \\
\end{center}
\caption{
\label{specplot1}
Sections of the STIS/E230H spectra of \hdone\
around lines of interest.
The filled squares represent the observed spectrum.
The solid blue line represents a synthetic spectrum with the 
best-fit abundance for each line of interest,
and the light-blue bands represent a change in this abundance
by a factor of $\pm$~2 (0.3~dex).
Blue lines without this light blue band denote upper limits.
The solid black line represents a synthetic spectrum 
with no contributions from the species of interest.
}
\end{figure*}

\begin{figure*}
\begin{center}
\includegraphics[angle=0,width=3.35in]{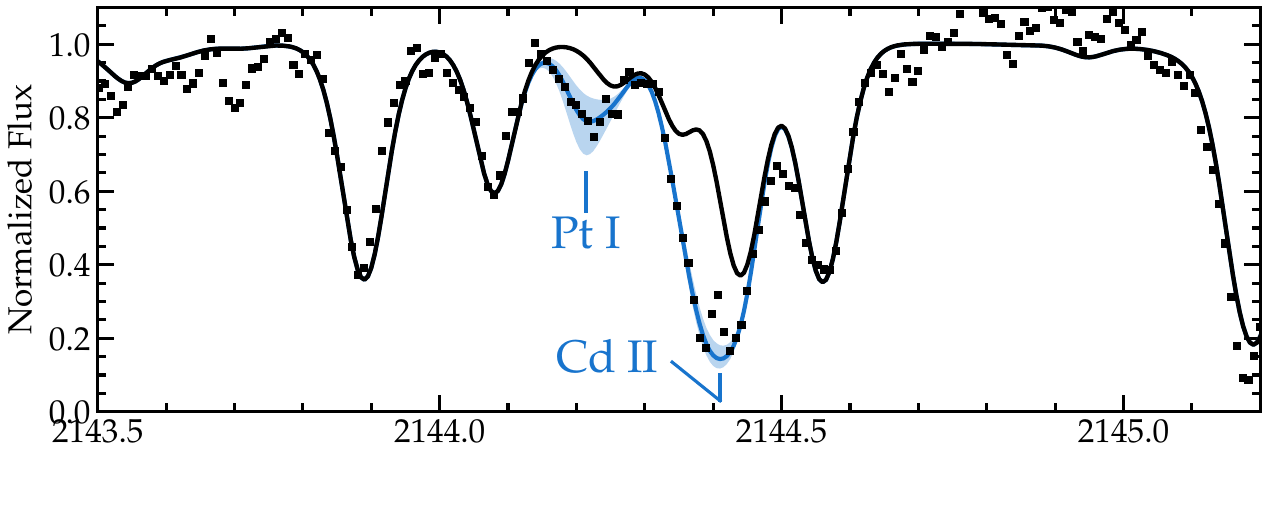}
\hspace*{0.1in}
\includegraphics[angle=0,width=3.35in]{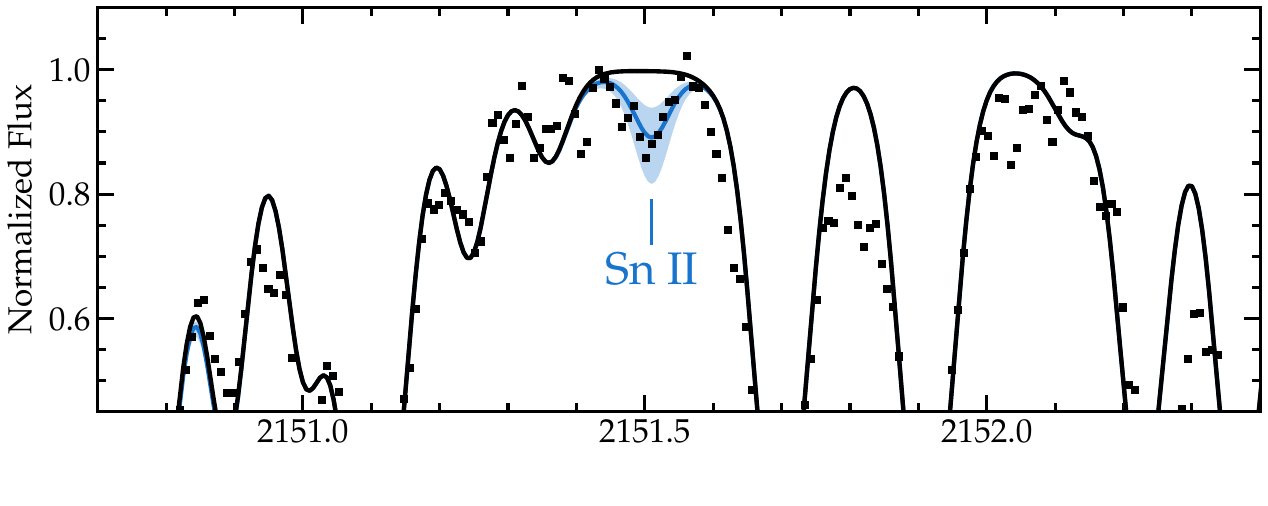} \\
\vspace*{-0.1in}
\includegraphics[angle=0,width=3.35in]{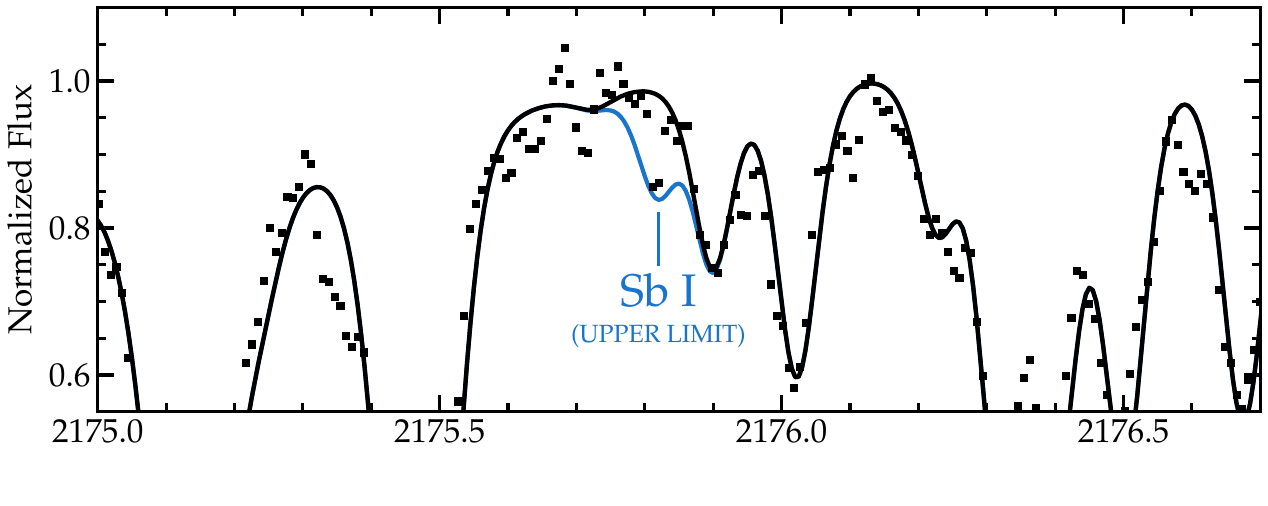}
\hspace*{0.1in}
\includegraphics[angle=0,width=3.35in]{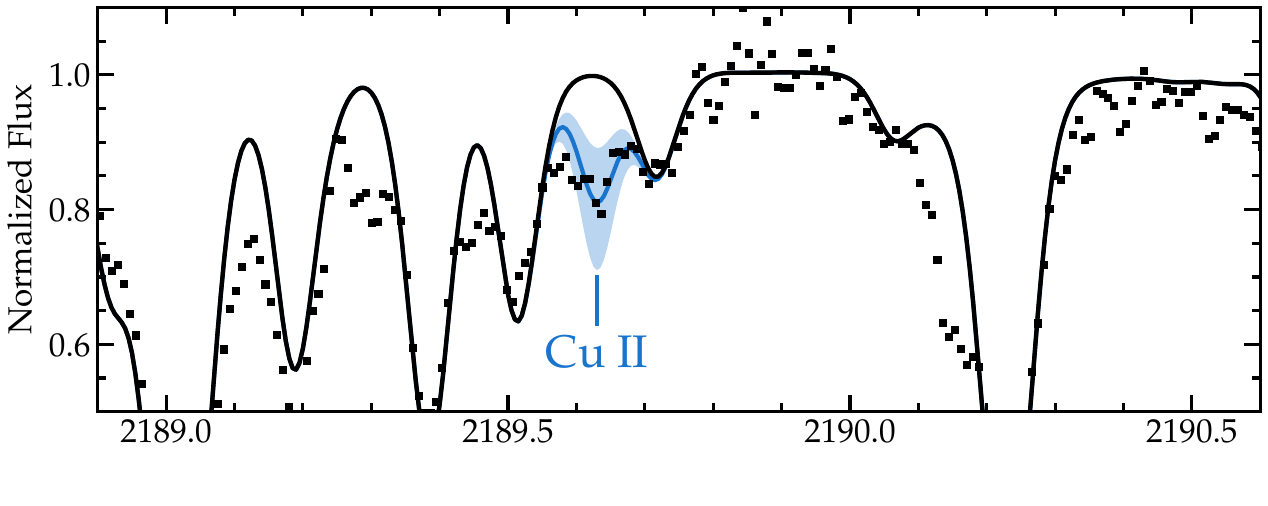} \\
\vspace*{-0.1in}
\includegraphics[angle=0,width=3.35in]{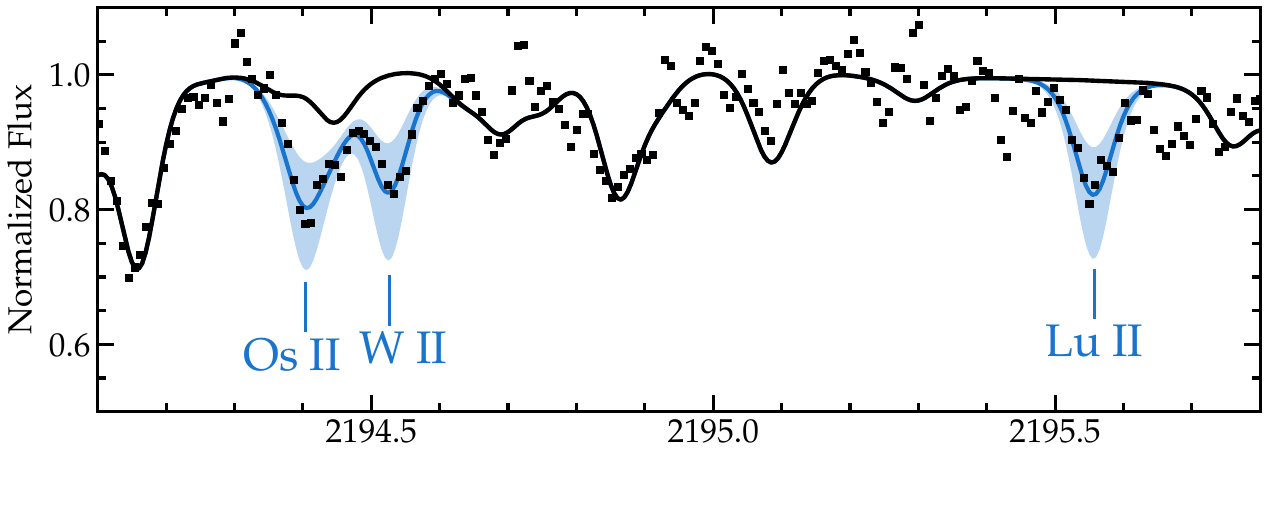}
\hspace*{0.1in}
\includegraphics[angle=0,width=3.35in]{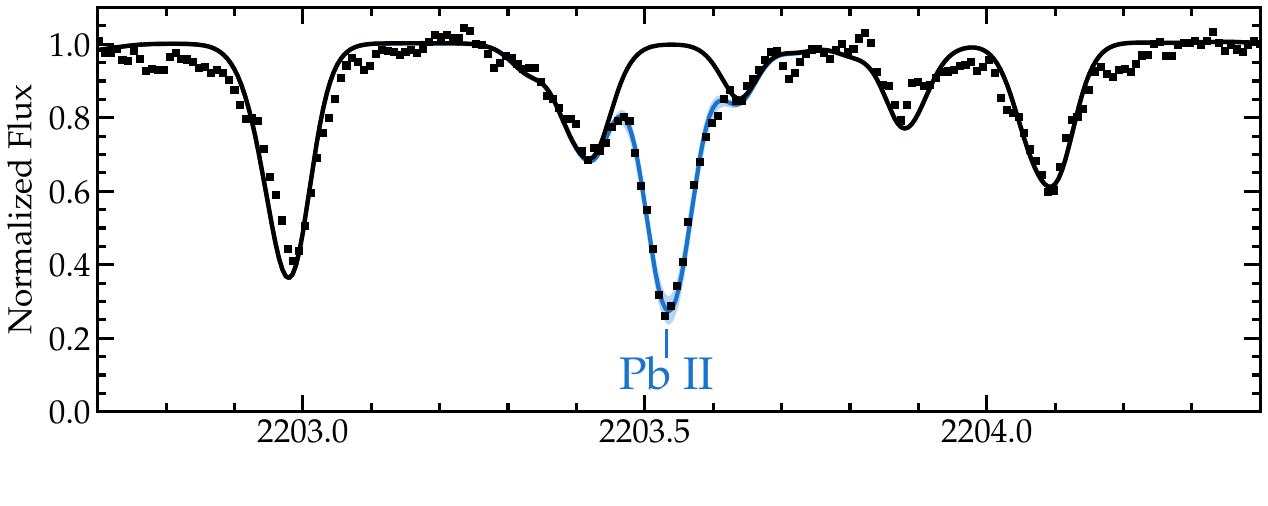} \\
\vspace*{-0.1in}
\includegraphics[angle=0,width=3.35in]{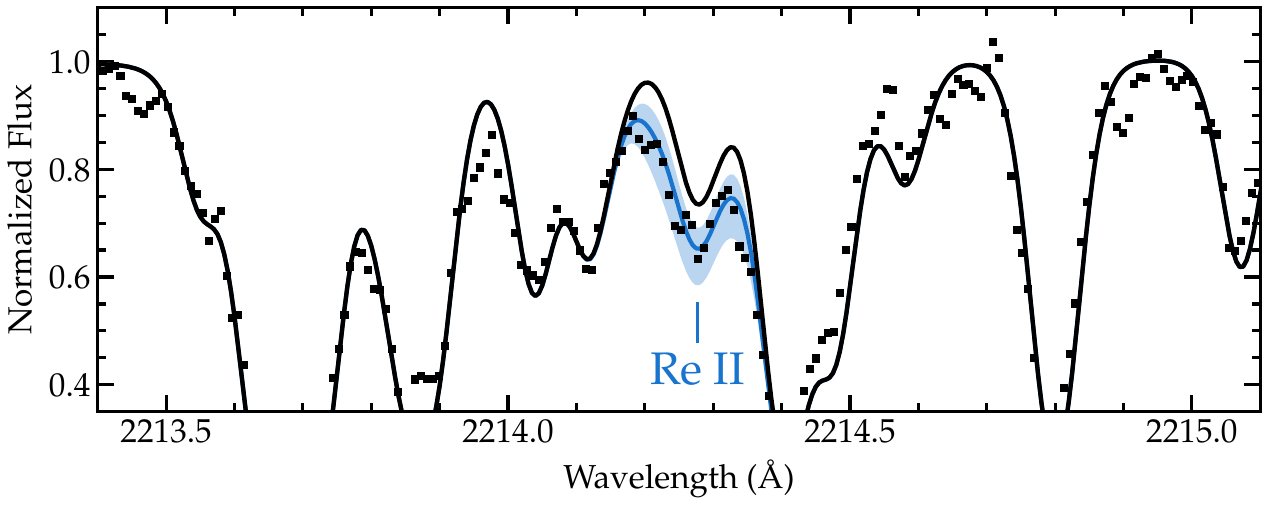}
\hspace*{0.1in}
\includegraphics[angle=0,width=3.35in]{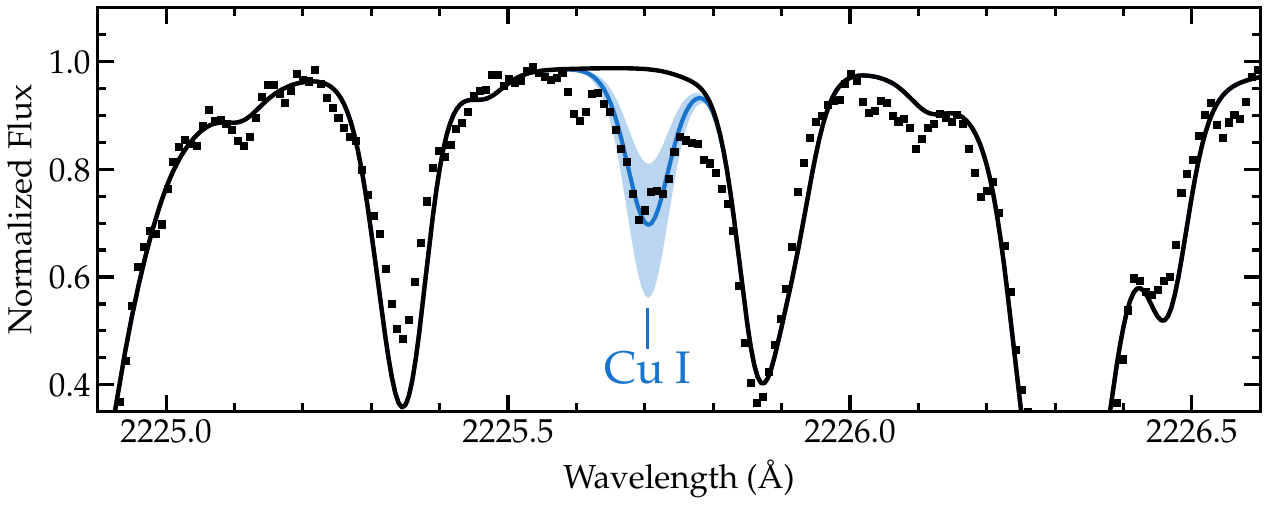} \\
\end{center}
\caption{
\label{specplot2}
Sections of the STIS/E230H spectra of \hdone\
around lines of interest.
Symbols are the same as in Figure~\ref{specplot1}.
}
\end{figure*}

\begin{figure*}
\begin{center}
\includegraphics[angle=0,width=3.35in]{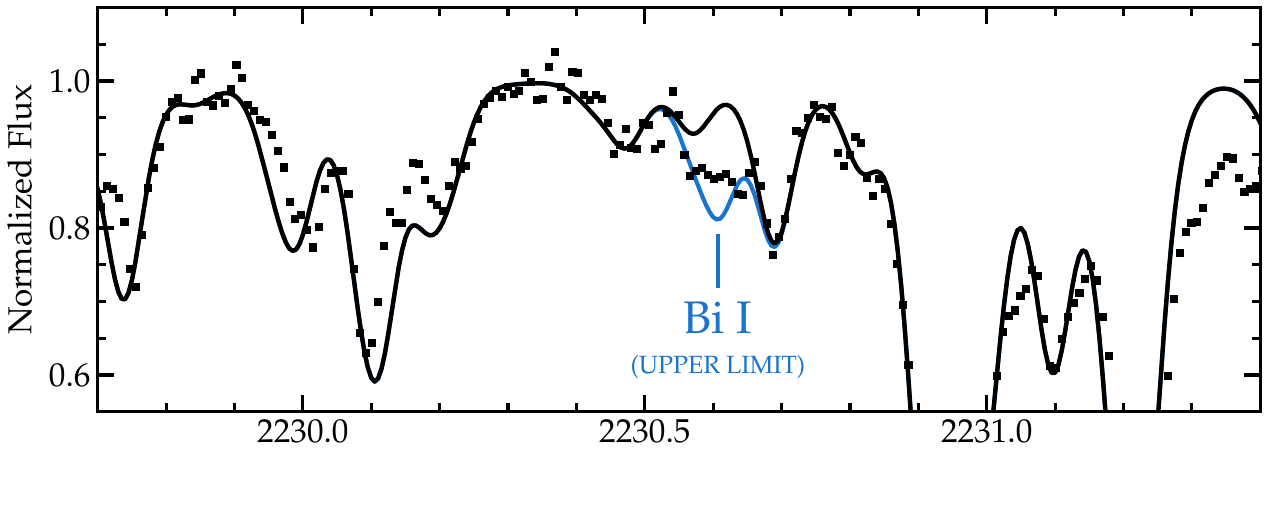}
\hspace*{0.1in}
\includegraphics[angle=0,width=3.35in]{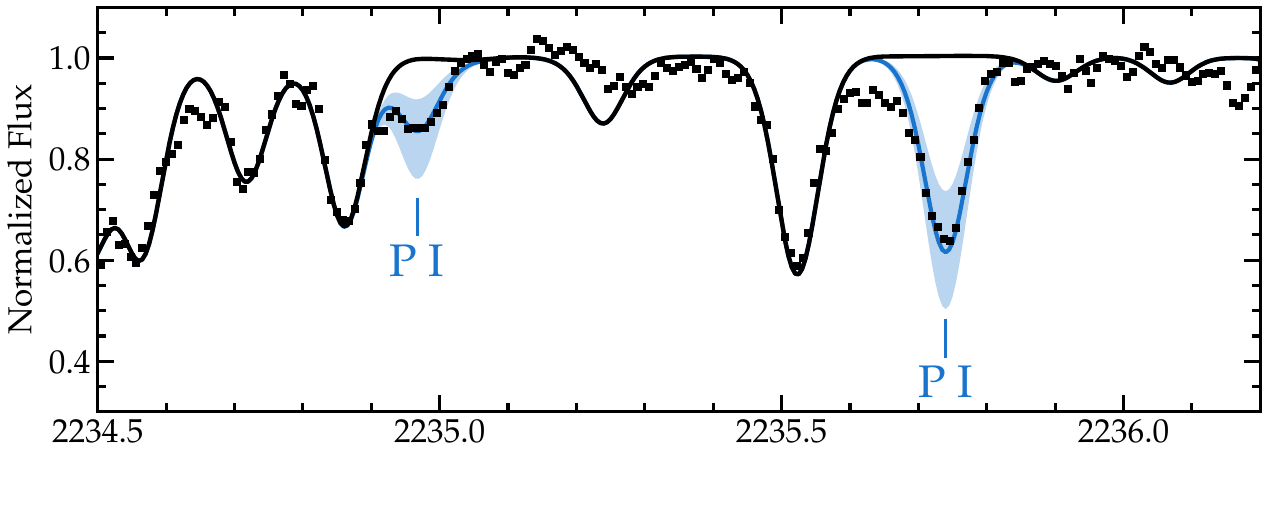} \\
\vspace*{-0.1in}
\includegraphics[angle=0,width=3.35in]{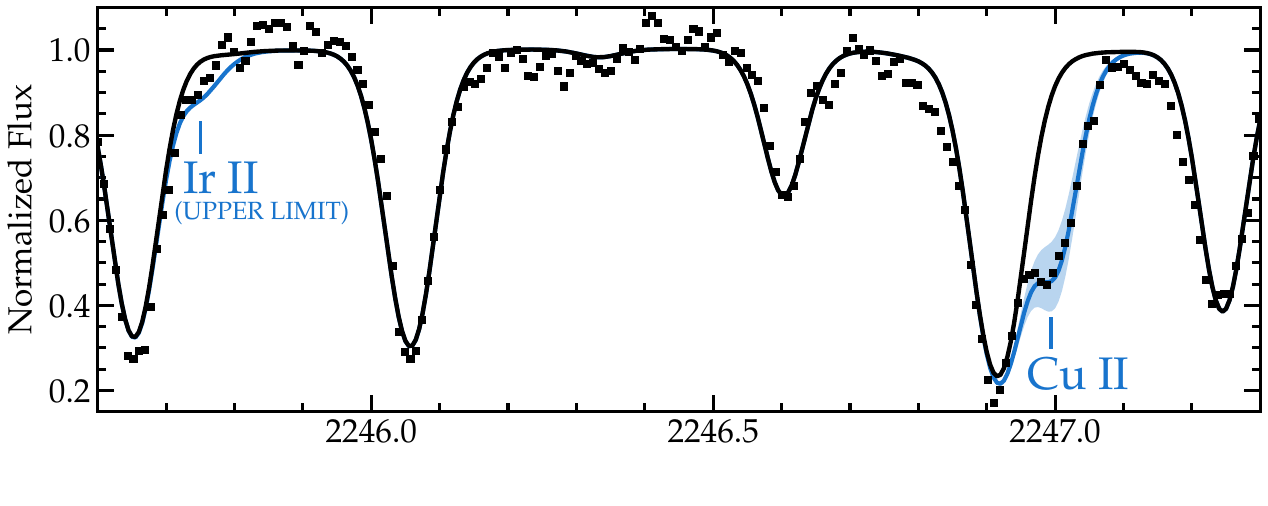}
\hspace*{0.1in}
\includegraphics[angle=0,width=3.35in]{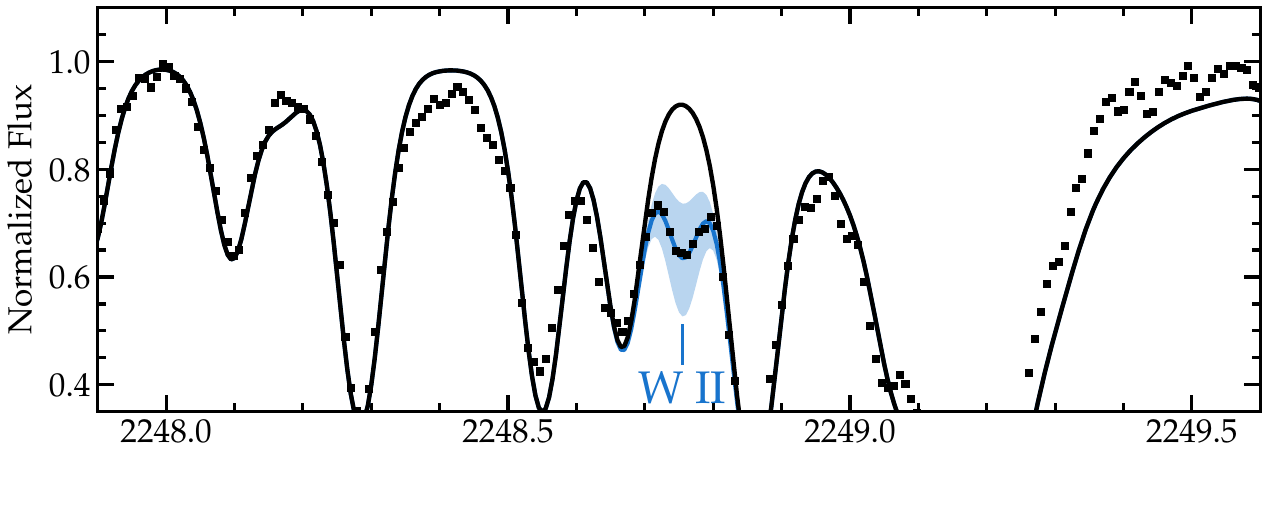} \\
\vspace*{-0.1in}
\includegraphics[angle=0,width=3.35in]{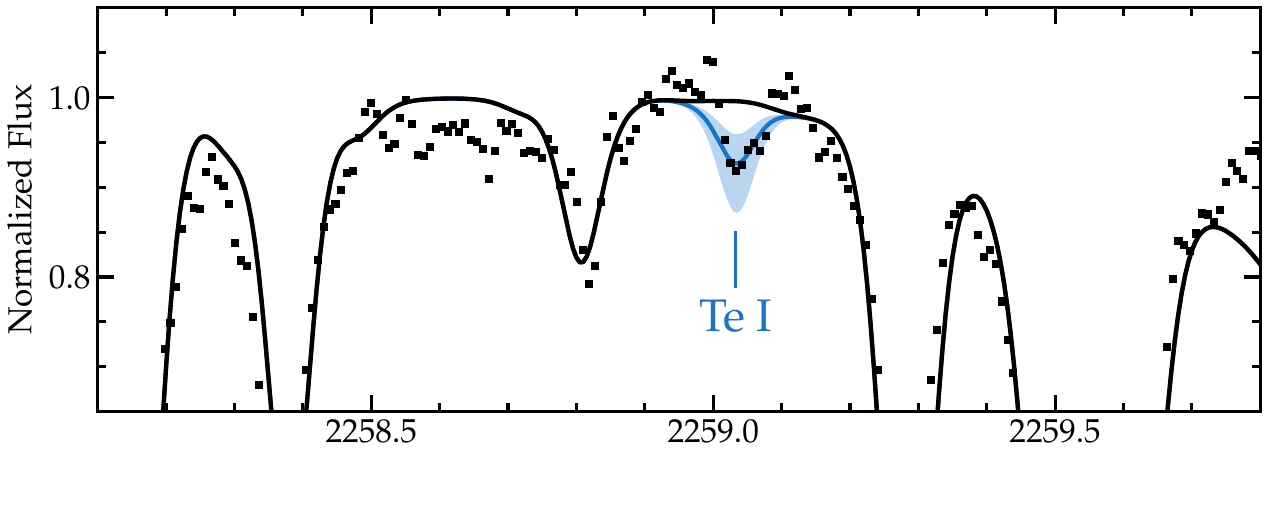}
\hspace*{0.1in}
\includegraphics[angle=0,width=3.35in]{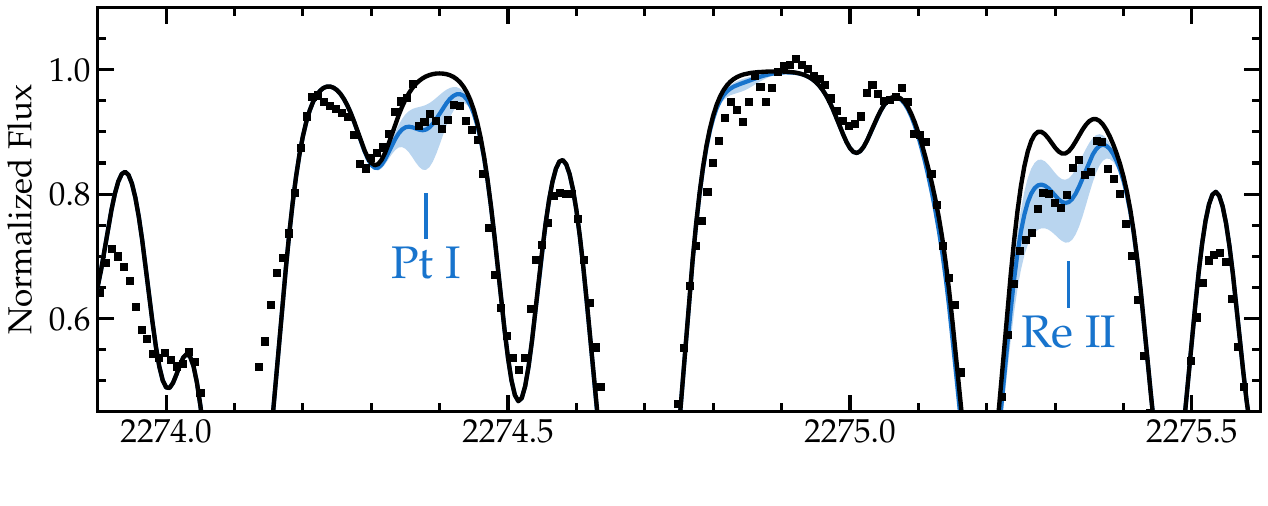} \\
\vspace*{-0.1in}
\includegraphics[angle=0,width=3.35in]{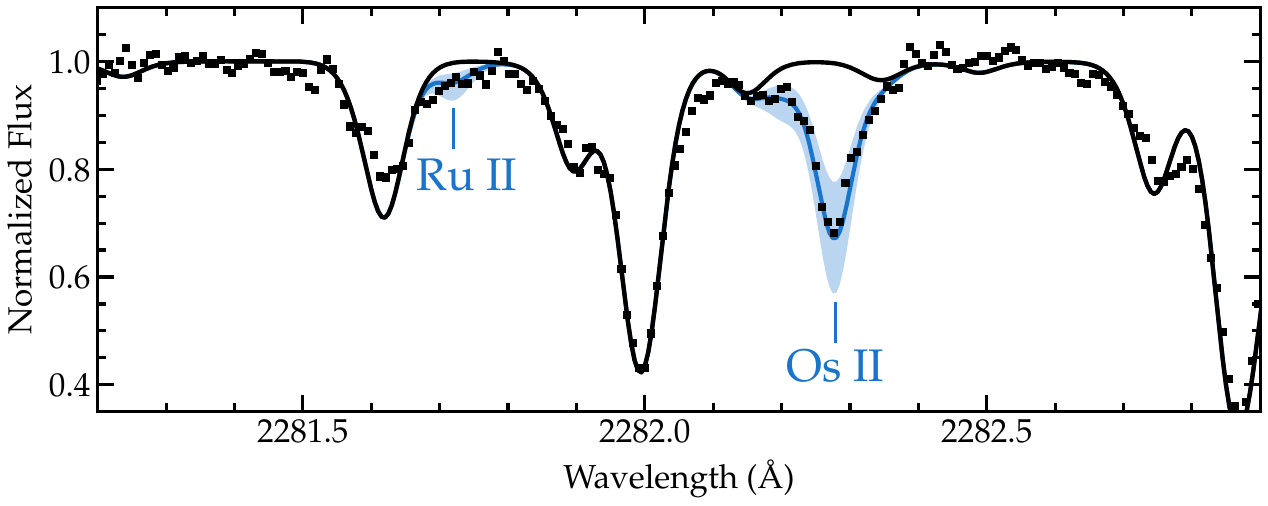}
\hspace*{0.1in}
\includegraphics[angle=0,width=3.35in]{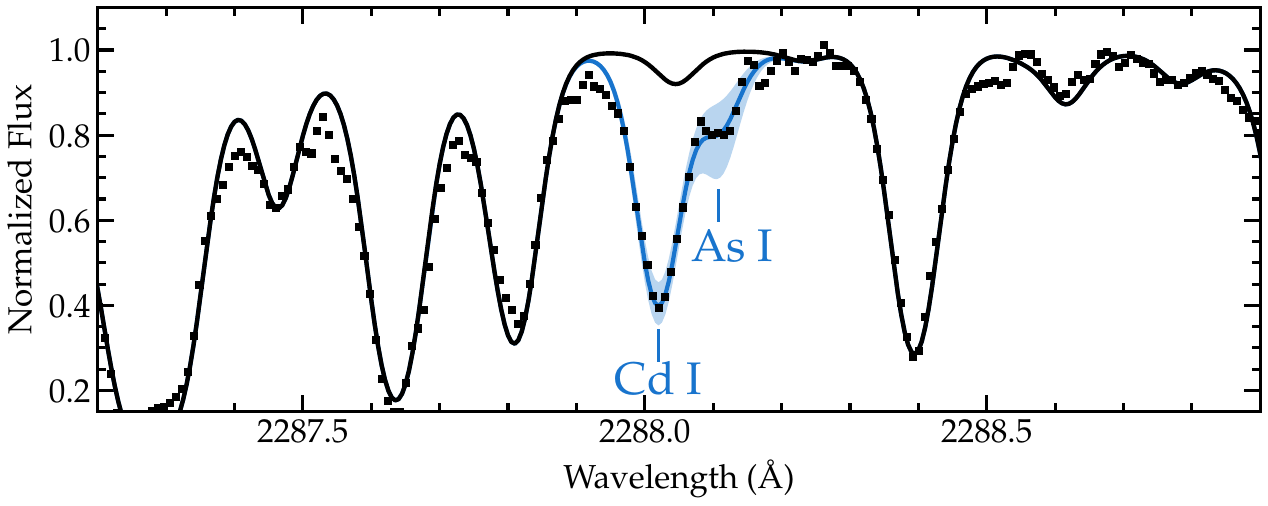} \\
\end{center}
\caption{
\label{specplot3}
Sections of the STIS/E230H spectra of \hdone\
around lines of interest.
Symbols are the same as in Figure~\ref{specplot1}.
}
\end{figure*}

Figures~\ref{specplot1}--\ref{specplot3} illustrate regions of the
STIS/E230H spectrum around important lines.
Several upper limits are illustrated, too.
We discuss these features individually
in Appendix~\ref{appendix}.

When matching the synthetic spectra to the observed spectrum,
we adjust the \loggf\ values of 
any line near the line of interest
if the \loggf\ value of the neighboring line
has not been determined by laboratory measurements
(e.g., \citealt{obrian91}) or 
theoretically-based empirical analysis (e.g., \citealt{peterson17}).
We also adjust the wavelengths of neighboring lines by up to 0.015~\AA\ 
to match the observed line profiles, although in practice
such shifts are rare and usually much less than this value.
Finally, whenever we identify no plausible transition
in the lists cited in Section~\ref{felinelist}
that can account
for absorption at a wavelength that blends with the wings
of a line of interest, we create an artificial Fe~\textsc{i} line with
E.P.\ = 0.0~eV whose strength is adjusted to match the 
observed line profile.
We note these cases in Appendix~\ref{appendix}.

Several of the elements of interest
have multiple stable isotopes.
We account for 
the small IS in the wavelengths
and HFS structure
whenever these data are available.
For B and Cu, we adopt the Solar isotopic abundance ratios
$^{11}$B/$^{10}$B $=$~4.045 and
$^{63}$Cu/$^{65}$Cu $=$~2.247 from \citet{meija16}.
For all heavier elements for which these data are 
available---Cd, Sb, Yb, W, Re, Ir, Pt, and Pb---we adopt the
\spro\ isotopic abundance fractions from \citet{sneden08}.

Table~\ref{abundtab} lists the weighted mean abundances 
derived from our study.
\citeauthor{placco15cemps}\ did not list 
uncertainties for abundances derived from individual lines,
so we adopt a typical uncertainty of 0.20~dex
for each line in their study.
We adopt the Hf and Pb abundances directly
from \citet{denhartog21hf} and \citet{roederer20}, respectively.

\input{tab2}

\subsection{NLTE Corrections}
\label{sec:nlte}

Non-LTE (NLTE) calculations relevant to the conditions found in
late-type metal-poor stellar atmospheres
are not available for most of the elements examined.
Here we discuss the few elements for which 
NLTE calculations have been performed,
and we estimate the potential impact of NLTE corrections to 
the LTE abundances.

\citet{placco15cemps} used two Sr~\textsc{ii} lines,
at 4077 and 4215~\AA, to derive
the Sr abundance in \hdone.
According to the NLTE calculations performed by \citet{bergemann12sr},
\hdone\ lies in a region of parameter space where the NLTE corrections
to these two lines are minimal, 
$\approx -$0.01~dex.
Thus the LTE abundance for Sr in \hdone\ is reasonably reliable.

\citet{placco15cemps} used five Y~\textsc{ii} lines
to derive the Y abundance in \hdone.
These lines originate from levels with moderate excitation potentials
($\approx$~1~eV).
The calculations of \citet{storm23} predict relatively small
NLTE corrections, $<$~0.05~dex or so,
to the LTE abundances derived from these lines
for stars similar to \hdone.
The LTE Y abundances are therefore likely to be
nearly correct.

\citet{placco15cemps} used three Ba~\textsc{ii} lines to derive the
Ba abundance in \hdone.
\citet{mashonkina14baeu} presented NLTE corrections for Ba abundances
for these three lines in a limited selection of model atmosphere
parameter combinations.
None of those models directly matches the parameters for \hdone.
Five of the models that are close to our model for \hdone\
predict mean NLTE corrections for these lines
in the range of $\approx -$0.20 to $\approx -$0.10~dex.

\citet{mashonkina14baeu} also presented NLTE corrections for Eu abundances
derived from the three lines of Eu~\textsc{ii} that we use to 
rederive the Eu abundance in \hdone\ (Section~\ref{sec:eu}).
As for Ba, only a limited selection of model atmosphere parameter combinations
are available.
Four of the models with similar sets of parameters
predict mean NLTE corrections for these three lines
of $+$0.06, $+$0.07, $+$0.12, and $+$0.14~dex,
an average of $+$0.10~dex.
Similar results are obtained from the calculations of \citet{guo25}.
Thus our derived LTE Eu abundance could 
slightly underestimate the Eu abundance.

As discussed in Appendix~\ref{appendix},
LTE calculations for Ge and Bi are likely susceptible to
NLTE overionization, and 
their abundances or upper limits
presented in Table~\ref{abundtab}
might be underestimated by a few tenths of a dex.
We encourage new investigations of NLTE calculations for these species.

\section{Results}
\label{results}

Our recommended neutron-capture abundances in \hdone\
are listed in Table~\ref{finaltab}.
These abundances are compiled from
our results and several other sources, 
with references listed in Table~\ref{finaltab}.
The abundances of lighter elements in \hdone\ can be found
in Table~\ref{abundtab},
the references listed in Table~\ref{finaltab}, 
and \citet{roederer21}.

\input{tab3}

A total of 35~elements heavier than Zn
have been detected 
(Ga, Ge, As, Se, Sr, Y, Zr, Nb, Mo, Ru, 
Pd, Cd, Sn, Te, Ba, La, Ce, Pr, Nd, Sm, 
Eu, Gd, Dy, Ho, Er, Tm, Yb, Lu, Hf, W, 
Re, Os, Pt, Au, and Pb),
and upper limits are available for
nine other neutron-capture elements
(Rb, Rh, Ag, In, Sb, Tb, Ir, Bi, and Th).

\begin{figure*}
\begin{center}
\includegraphics[angle=0,width=5.5in]{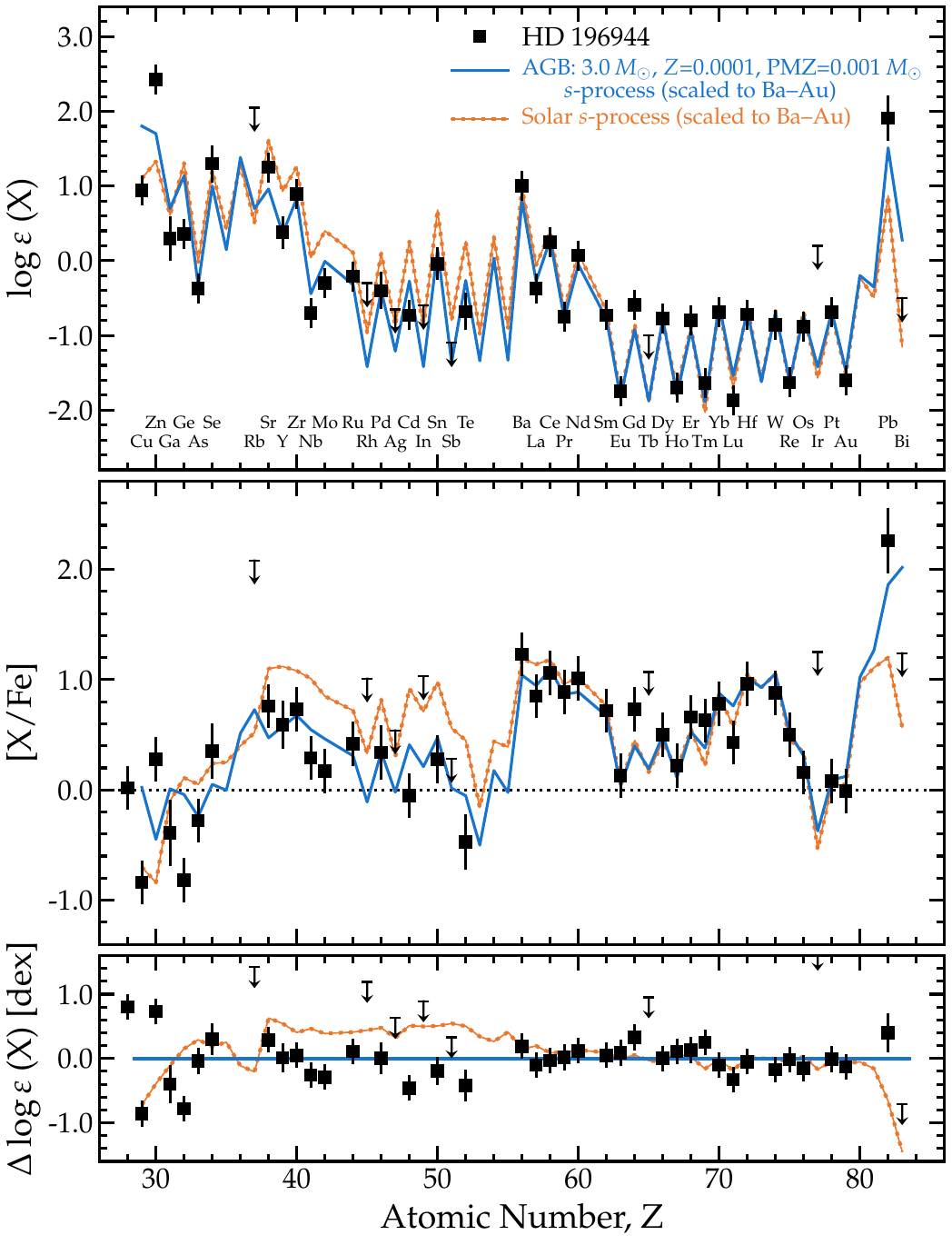}
\end{center}
\caption{
\label{abundplot}
Comparison between the \hdone\ abundance pattern
and model predictions for the 
\spro\ operating in an AGB star with initial mass = 3.0~\msun,
$Z$ = 0.0001, and partial mixing zone (PMZ) = 0.001~\msun\
\citep{karakas10,lugaro12}.
The Solar \spro\ pattern from \citet{bisterzo11} (for Ga--Br)
and \citet{bisterzo14} (for Rb--Bi) are shown in orange studded lines
for comparison.
The comparison is shown in terms of $\log\varepsilon$ abundances (top),
[X/Fe] ratios (middle), and differences between the \hdone\ pattern and the
predicted \spro\ pattern (bottom).
}
\end{figure*}

\subsection{Sensitivity of Derived Abundances to the
Choice of Model Atmosphere}
\label{sec:modelcomp}

Spectroscopically derived model atmospheres have long been known
to be cooler and yield lower surface gravity values than 
photometrically derived ones
(e.g., \citealt{johnson02abund,frebel13,jofre19}).
\citet{placco15cemps} derived the model atmosphere parameters 
for \hdone\ spectroscopically.
\citet{mittal25} derived \teff\ using
photometric color-\teff-metallicity relations, and
\logg\ using fundamental physical relationships that
rely on the stellar parallax.
\citeauthor{mittal25} derived a model atmosphere, interpolated
from the MARCS grid \citep{gustafsson08}, with
\teff\ = 5428 $\pm$~29~K,
\logg\ = 2.11 $\pm$~0.16,
\vt\ = 1.70 $\pm$~0.06~\kmsec, and
[Fe/H] = $-$2.09 $\pm$~0.07.
These quoted uncertainties represent statistical ones, and
[Fe/H] is based on NLTE corrections to 
LTE abundances derived from Fe~\textsc{i} lines.
This warmer model atmosphere
increases the abundances 
for all metals.
We adopt the cooler, spectroscopically derived model for our analysis
for consistency with previous work.
We assess the impact of this choice 
by deriving abundances using both model atmospheres
for a few key neutron-capture elements.

The mean [Se/Fe], [Te/Fe], and [Pt/Fe] abundance ratios,
which are derived from lines of neutral atoms,
are lower for the cooler model by $-$0.33 $\pm$~0.04 ($\sigma = 0.06$) dex.
The mean [Ba/Fe], [Eu/Fe], and [Pb/Fe] abundance ratios,
which are derived from lines of singly ionized atoms,
are lower for the cooler model by $-$0.07 $\pm$~0.05 ($\sigma = 0.09$) dex.
These differences are 
comparable to the scale of the abundance uncertainties.
The differences among ratios
of neutron-capture elements are smaller when comparing
abundances derived from like ionization states.
The [Te/Se], [Pt/Te], [Ba/Eu], and [Pb/Ba] ratios are 
higher for the cooler model
by only $+$0.11 $\pm$~0.03 ($\sigma = 0.02$) dex.
We suggest that this difference represents the minimum systematic uncertainty.

In cases where abundance ratios must be constructed from
mixed ionization states, such as 
[Se/Ba], [Te/Ba], or [Pt/Ba], 
the differences are lower for the cooler model 
by $-$0.25 $\pm$~0.04 ($\sigma = 0.06$) dex.
We suggest that this range,
$\approx -0.25$ to $+$0.1~dex,
provides an estimate for the systematic uncertainty in the
neutron-capture abundance ratios.
Ratios derived from lines of the same ionization state
minimize the systematic uncertainty, 
while 
ratios derived from lines of different ionization states
are more sensitive to the choice of model atmosphere parameters.

\section{Discussion}
\label{discussion}

\subsection{AGB Model Fits}
\label{sec:agbmodel}

Figure~\ref{abundplot}
illustrates the heavy-element abundances in \hdone.
The blue line in Figure~\ref{abundplot} marks
the \spro\ elemental abundances produced 
by a model of an AGB star with
initial mass of 3~\msun\ and $Z$ = 0.0001 ([Fe/H] = $-$2.2).
The orange line marks the Solar \spro\ pattern.
Both the AGB model predictions and the Solar pattern
have been scaled down to match the approximate 
abundance level in \hdone\ for
56~$\leq Z \leq$~79 (Ba--Au).

The blue line shown in Figure~\ref{abundplot} 
represents the overall
best fit to the observed abundance pattern
at the end of the thermally pulsing stage of an AGB star.
This fit is considerably better than the Solar \spro\ pattern.
The AGB predictions were originally made by \citet{lugaro12}
using the stellar evolutionary sequences from \citet{karakas10}. 
\citeauthor{lugaro12}\ calculated a series 
of low-metallicity AGB nucleosynthesis models 
across a range of stellar masses 
(1 to 6~\msun), 
testing different initial compositions in order to study 
the origins of \cemps\ and \cemprs\ stars. 
While the focus of that paper was on the \spro\ operating at low metallicity, 
the calculations included a full nuclear network from H to Po
(1 $\leq Z \leq$ 84).
The 3~\msun\ model is hot enough for the
$^{22}$Ne($\alpha$,$n$)$^{25}$Mg
source to operate during convective thermal pulses, 
alongside the 
$^{13}$C($\alpha$,$n$)$^{16}$O reaction,
which produces neutrons during the interpulse phase,
as a result of the $^{13}$C pockets
(see review by \citealt{lugaro23}).
This scenario yields increased production of 
elements near the first \spro\ peak
(Rb, Sr, Y, Zr)
relative to AGB model predictions of lower mass for the same metallicity.

The AGB model prediction provides a remarkably good fit to the observed
abundance pattern
when diluted by a factor of $\approx$~1~dex.
The differences between the observed and predicted abundances are
no larger than $\approx$~2 times the observational uncertainty
for all elements with 33 $\leq Z \leq$~82,
as shown in the bottom panel of Figure~\ref{abundplot}.
When scaled as shown in Figure~\ref{abundplot}, some elements
at the \spro\ peaks are slightly underpredicted by this model
(Sr, Ba, Pb), while others are not
(Y, Zr, La, Ce, Pr, Nd).
We checked whether these differences could be further minimized
by truncating the accretion of freshly produced \spro\ material
onto \hdone\ at a thermal pulse earlier than the final one (pulse 20).
The ratios of the [Ba/X] abundances, where
X denotes an element between the second and third \spro\ peaks,
change by $<$~0.2~dex from thermal pulse 5 to 20.
This test demonstrates that it is not possible to 
further improve the already satisfactory fit.
This overall agreement is noteworthy because
several of the heavy elements detected in the UV spectrum of \hdone\
have only been studied 
in the context of an \spro\ environment 
in Solar System material
or post-AGB stars (e.g., \citealt{vanaarle13,desmedt16}).

It remains challenging, however, to constrain
the characteristics of the former AGB companion to \hdone\
with great precision.
Fitting the abundance pattern, and sometimes additional
properties of the system, such as the orbital period, has yielded 
predictions for the initial mass of that AGB companion 
of 0.9~\msun, 1.4~\msun, 1.5~\msun, and 3.0~\msun\
(\citealt{placco15cemps}, \citealt{abate15}, \citealt{bisterzo11}, and 
this study, respectively).
Exploring the differences between these models and approaches
is beyond the scope of the present study.
We simply conclude that
there is general agreement that the companion
was a low-mass star that passed through the AGB phase of evolution,
and neutrons for the \spro\ were likely produced
by both the 
$^{13}$C($\alpha$,$n$)$^{16}$O and 
$^{22}$Ne($\alpha$,$n$)$^{25}$Mg reactions.

\subsection{Minimal \textit{s}-process Contribution up to Selenium}

Figure~\ref{abundplot} shows a poor match between the 
AGB model predictions and the observational data
for the elements at the heavy end of the iron group
and just beyond:\ Ni, Cu, Zn, Ga, and Ge (28 $\leq Z \leq$ 32).
These five elements, and the next two heaviest ones, As and Se
(33 $\leq Z \leq$ 34),
have only been detected together in one other metal-poor star, 
\hdtwo\ \citep{roederer22a},
which is highly enhanced in \rpro\ elements.
\citeauthor{roederer22a}\ concluded from their analysis of the
abundance pattern in \hdtwo\ that 
Se was the lightest element with a substantial \rpro\ contribution,
and the \rpro\ produced no more than minimal amounts of Ga, Ge, or As.

We compare the ratios of the abundances of
Ga, Ge, and As in \hdone\ and \hdtwo.
These three elements are found in similar proportions in both stars.
The \logeps{Ga/Ge} and \logeps{As/Ge} ratios are
$-$0.06 and $-$0.73, respectively, in \hdone, 
and these ratios are $-$0.20 and $-$0.45, respectively, in \hdtwo.
All three elements exhibit similar levels of enhancement relative to Fe
in both stars, too.
The [Ga/Fe], [Ge/Fe], and [As/Fe] abundances are
$-$0.39, $-$0.82, and $-$0.28, respectively, in \hdone, and 
these three ratios are 
$-$0.32, $-$0.73, and $+$0.17, respectively, in \hdtwo.
The ratios among Ga, Ge, and Fe are 
virtually indistinguishable in the two stars,
and As is---at most---only slightly enhanced in \hdtwo.

\citet{roederer22a} concluded that the \rpro\ did not 
contribute substantially to the Ga, Ge, or As in \hdtwo.
The similarity of these ratios in both
\hdone\ (an \spro-enhanced star)
and \hdtwo\ (an \rpro-enhanced star)
lead us to conclude that this 
\spro\ also did not contribute substantially
to Ga, Ge, or As.
This result is a consequence of the high neutron/Fe ratios
found in both 
the \spro\ operating in a low-metallicity environment and the \rpro,
wherein each Fe seed nucleus captures many neutrons
and flows through relatively light nuclei to heavier ones.
This finding echos that of \citet{roederer12c} and \citet{roederer14d},
who found a constant [As/Fe] ratio in seven stars with 
$-$2.8 $\leq$ [Fe/H] $\leq -$0.8.

We consider whether the intermediate neutron-capture process
(\ipro) operating in the companion AGB star
could have potentially been responsible for the
production of these three elements.
We have checked the predicted \ipro\ abundance ratios for 
models\footnote{\href{http://www.astro.ulb.ac.be/~siess/Site/StellarModels}{http://www.astro.ulb.ac.be/{$\sim$}siess/Site/StellarModels}}
by \citet{choplin22,choplin24},
with initial masses spanning 1 to 3~\msun, 
metallicities spanning [Fe/H] from $-$3 to $-$2, and 
with and without overshooting.
No model in this grid offers a satisfactory explanation for
the abundance pattern of \hdone.
When scaled to the abundances of Ga, Ge, and As,
most of these models overproduce all other heavier elements,
typically by factors $\geq$~3.
The 3~\msun\ models with no overshooting
produce [X/As] ratios comparable to the \spro\ model predictions
for elements, X, between the first and second \spro\ peaks.
Including this \ipro\ material, however,
would result in considerably worse fits in Figure~\ref{abundplot}
when added to the \spro\ material required to reproduce
the abundances of heavier elements.
It is also difficult to envision a scenario where \hdone, \hdtwo, and
a fair fraction of other metal-poor stars
were all enriched with AGB \ipro\ material containing
similar proportions of Ga, Ge, and As.
We propose that these elements were synthesized
by some other process(es) common to both sets of
massive-star supernovae that enriched the gas
from which \hdone\ and \hdtwo\ formed.

\subsection{Non-detection of Bismuth}
\label{sec:bi}

Bismuth is undetected in our spectrum of \hdone,
as shown in Figure~\ref{specplot3} and discussed in 
Appendix~\ref{bismuth}.
The upper limit derived from the Bi~\textsc{i} line at 2230~\AA, 
[Bi/Fe] $< +$1.24, 
lies more than 0.7~dex below the abundance predicted by our
best-fit AGB model.
The neighboring third-peak element Pb is also considerably 
more abundant than Bi, [Pb/Bi] $> +$1.02.

Several sets of nucleosynthesis models predict similar ratios 
among the Pb and Bi abundances.
Our best-fit model predicts [Pb/Bi] = $-$0.16.
The analogous model
(3~\msun, $Z$ = 0.0001)
in the
FUll-Network Repository of Updated Isotopic Tables \& Yields
(FRUITY) database \citep{cristallo11} 
predicts a similar [Pb/Bi] ratio, $+$0.19.
There is no reliable analogous model
in the NuGrid database,
but the 3~\msun\ $Z$ = 0.001 model from \citet{battino21}
also predicts a similar [Pb/Bi] ratio, $+$0.29.
These predictions are in fair agreement,
and we conclude that the predicted [Pb/Bi] ratio is relatively
insensitive to the choice of the reaction network
and stellar model.

One potential explanation for this discrepancy is that the
Bi~\textsc{i} line is not formed in LTE.~
NLTE overionization may underestimate abundances
derived from lines of the neutral species when that species is
a significant minority population.
Overionization of neutral Bi is likely to occur,
because the first ionization potential (FIP) of Bi is 7.29~eV,
and so neutral Bi is a minority species in the atmosphere of \hdone.
No NLTE calculations are available for Bi in metal-poor stars.
If the predicted Bi abundance is correct, 
we would expect that the NLTE correction to the Bi abundance
must be at least $+$0.7~dex.
New NLTE calculations for Bi line formation, and additional
observations of Bi lines in metal-poor stars,
would help to understand this discrepancy.

\subsection{The Classification of \mbox{HD~196944}}
\label{sec:eu}

The classification of \hdone\ as a \cemps\ or \cemprs\ star 
depends on the C, Ba, Eu, and Fe abundances \citep{beers05}.
The carbon enhancement is 
[C/Fe] = $+$1.31 $\pm$~0.20, 
after accounting for evolutionary effects on the C abundance \citep{placco14c},
securing its status as a CEMP star.

The Eu abundance in \hdone\ has been difficult to establish.
The difficulty likely arises from a combination of factors, including
its relatively low abundance, weak lines,
and placement of its strongest lines in the blue spectral region
near 4000~\AA\ where line blending and continuum placement
are challenging because of the strong CH and CN features in CEMP stars.
Previous LTE abundance determinations 
of the [Eu/Fe] ratio in \hdone\ include those by
\citet{aoki02pb} ($+$0.17 $\pm$ 0.19),
\citet{masseron10} ($+$0.22 $\pm$ 0.20),
\citet{roederer14c} ($+$0.12 $\pm$ 0.09),
\citet{placco15cemps} ($-$0.11 $\pm$ 0.10), and
\citet{karinkuzhi21} ($+$0.50 $\pm$ 0.20).

We rederive the Eu abundance using three sets of model atmosphere parameters:\
\teff/\logg/[Fe/H]/\vt\ = 
5168~K/1.28/$-$2.50/1.68~\kmsec\ \citep{karinkuzhi21},
5170~K/1.60/$-$2.41/1.55~\kmsec\ \citep{placco15cemps}, and
5428~K/2.11/$-$2.10/1.70~\kmsec\ \citep{mittal25}.
These parameters span the range adopted by previous studies.
Three Eu~\textsc{ii} lines are available to us 
in the MIKE and 2dCoud\'{e} spectra described in 
Section~\ref{sec:previousspec},
at 3819, 4129, and 4205~\AA.~
For the first (coolest) model, we derive $\log\varepsilon$ abundances
from these three lines of $-$1.95, $-$1.92, and $-$1.69, respectively,
and a mean abundance of $-$1.85
([Eu/Fe] = $+$0.13).
The next model is the one we have adopted throughout this study.
For this model, we derive $\log\varepsilon$ abundances 
from these three lines of $-$1.86, $-$1.80, and $-$1.62, respectively,
for a mean of $-$1.76
([Eu/Fe] = $+$0.13).
For the third (warmest) model, we derive $\log\varepsilon$ abundances
from these three lines of $-$1.52, $-$1.49, and $-$1.26, respectively,
and a mean abundance of $-$1.42
([Eu/Fe] = $+$0.16).
As discussed in Section~\ref{sec:nlte}, the NLTE corrections to the
LTE Eu abundance in \hdone\ are likely small, $\approx +$0.10~dex.
The [Eu/Fe] ratio for \hdone\ is $\approx +$0.15
in LTE, and $\approx +$0.25 if NLTE calculations are considered.

The LTE [Ba/Fe] ratio in \hdone\ is
$+$1.24 in LTE ($\approx +$1.1 in NLTE) \citep{placco15cemps}.
The LTE [Ba/Eu] ratio is thus 
$\approx +$1.1 ($\approx +$0.85 in NLTE).
We conclude that \hdone\ is best classified as
a \cemps\ star,
with [Ba/Fe] $> +$1.0 and [Ba/Eu] $> +$0.5.
This classification is further supported by the 
superb match between the observed abundances and the AGB \spro\ 
model predictions (Section~\ref{sec:agbmodel}),
which were generated without any need for a
neutron density higher than that of the traditional \spro\
(cf.\ \citealt{dardelet15}).

\section{Conclusions}
\label{conclusions}

We have extended previous work by \citet{placco15cemps},
who performed the first abundance analysis based on the
UV spectra of \cemps\ stars,
to even shorter UV wavelengths
in the brightest \cemps\ star, \hdone.
The spectral region from 2029 $\leq \lambda \leq$ 2306~\AA\
permits the detection of several species that are not 
otherwise detectable in longer-wavelength UV, optical, or near-infrared 
spectra of this star, including
Ga~\textsc{ii},
As~\textsc{i},
Se~\textsc{i},
Ru~\textsc{ii},
Cd~\textsc{ii},
Sn~\textsc{ii},
Te~\textsc{i},
Lu~\textsc{ii},
W~\textsc{ii},
Re~\textsc{ii},
Os~\textsc{ii},
and
Pb~\textsc{ii}.
Nearly all elements with stable isotopes between
the second and third \spro\ peaks have been detected,
as well as many between the first and second peaks.

We establish the classification of \hdone\ as a \cemps\ star
through its heavy-element abundance pattern, and
as defined through is [C/Fe], [Ba/Fe], and [Ba/Eu] ratios.
We show that these abundances can be well-fit
by a model of \spro\ nucleosynthesis in a
low-mass AGB star that subsequently
transferred these freshly produced elements to \hdone.
The abundance ratios among Ga, Ge, and As 
(31 $\leq Z \leq$ 33)
are indistinguishable from those
in the highly \rpro-enhanced star \hdtwo,
indicating that neither the \spro\ (nor \rpro)
contributed substantially to the abundances of these elements.

Finally, our study can serve as a guide to 
future high-resolution UV spectroscopy
of additional \spro-enhanced stars.
Expanding the sample further than the two \cemps\ stars
already observed with STIS requires
observing fainter stars,
which is difficult with the limited sensitivity of HST.~
The proposed Habitable Worlds Observatory 
(HWO)
offers the potential to extend this work
by collecting similar quality UV spectra
of virtually any \cemps\ star
that has been observed in the optical from the ground.
The discovery potential of HWO for this science area is thus immense.

\begin{acknowledgments}

We thank Shivani Shah for helpful comments on an earlier version
of the manuscript and Maria Lugaro and Bal\'{a}zs Sz\'{a}nyi
for insightful discussions.
We
acknowledge generous support 
provided by NASA through grant GO-14765 
from STScI, which is operated by AURA,
under NASA contract NAS5-26555.
We acknowledge support 
awarded by the U.S.\ National Science Foundation (NSF):\
grants 
PHY~14-30152 (Physics Frontier Center/JINA-CEE),
OISE~1927130
(International Research Network for Nuclear Astrophysics/IReNA),
and
AST~2205847 (I.U.R.).
I.U.R.\ acknowledges support from the NASA
Astrophysics Data Analysis Program, grant 80NSSC21K0627.
The work of V.M.P.\ is supported by NOIRLab, 
which is managed by AURA under a cooperative agreement with 
the NSF.~
T.C.B.\ acknowledges partial support from grant DE-SC0023128CeNAM
(the Center for Nuclear Astrophysics Across Messengers/CeNAM),
awarded by the U.S.\ Department of Energy, 
Office of Science, Office of Nuclear Physics.
This research has made use of NASA's
Astrophysics Data System Bibliographic Services;
the arXiv pre-print server operated by Cornell University;
the SIMBAD and VizieR
databases hosted by the
Strasbourg Astronomical Data Center;
the ASD hosted by NIST;
the MAST at STScI; 
and
NOIRLab IRAF, which is distributed by the 
Community Science and Data Center at NSF NOIRLab, 
which is managed by AURA under a cooperative agreement with the 
NSF.~

\end{acknowledgments}

\facility{HST (STIS)}

\software{IRAF \citep{tody86,tody93,fitzpatrick24},
LINEMAKE \citep{placco21linemake},
matplotlib \citep{hunter07},
MOOG \citep{sneden73,sobeckmoog},
numpy \citep{vanderwalt11},
scipy \citep{jones01}}

\appendix
\restartappendixnumbering

\section{Discussion of Individual Elements and Lines}
\label{appendix}

\subsection{\texorpdfstring{Boron (B, $Z = 5$)}{Boron (B, Z=5)}}
\label{boron}

The B~\textsc{i} resonance doublet at 2088.888 and 2089.570~\AA\
is not detected in our spectrum of \hdone.
We derive upper limits from each of these two lines, as shown in
Figure~\ref{specplot1}.
The bluer of the two lines is blended with two stronger lines.
The redder of the two lines is relatively unblended, yet not
obviously detected, and this line provides the stronger abundance limit.

\citet{kiselman96} presented NLTE calculations for the
B~\textsc{i} resonance doublet examined here.
Their NLTE corrections are substantial,
reaching $\approx +$0.9~dex for stars like \hdone\
because of photon pumping in the resonance lines
(see also \citealt{asplund05}).
Our boron upper limit is derived from resonance lines, 
so it may be substantially underestimated in LTE.~

\subsection{\texorpdfstring{Phosphorus (P, $Z = 15$)}{Phosphorus (P, Z=15)}}
\label{phosphorus}

Several P~\textsc{i} lines are observed in our spectrum of \hdone.
Most of these lines are not useful as abundance indicators.
The P~\textsc{i} line at 2136.182~\AA\ is clean but saturated;
the lines at 2222.586 and 2223.358~\AA\ are too weak;
and
P~\textsc{i} lines at 2135.465, 2149.145, 2152.940, 2154.080, 2154.113,
and 2242.543~\AA\ are too blended.
Only two P~\textsc{i} lines, at 2234.961 and 2235.732~\AA, are
unblended and unsaturated, 
as shown in Figure~\ref{specplot3}.
The NIST ASD rates their \loggf\ values as grade E
(uncertainties $\geq$~50\%, or 0.30~dex).
Uncertainties in the atomic data dominate the error budget,
yet we derive concordant abundances from these two lines.

The two lines used as abundance indicators arise from the
excited $3s^{2}3p^{3}$ $^{2}$D$^{\rm o}$ level
of neutral P.~
NLTE calculations are not yet available for 
these P lines in late-type stars.

\subsection{\texorpdfstring{Copper (Cu, $Z = 29$)}{Copper (Cu, Z=29)}}
\label{copper}

Several lines of Cu~\textsc{i} are detectable in our spectrum of \hdone,
and all arise from the ground state.
Two of these, at 2165.093 and 2225.697~\AA, yield reliable abundances;
the $\lambda$2225 line is shown in Figure~\ref{specplot2}.
Two Cu~\textsc{i} lines at 2178.944 and 2181.720~\AA\ are too blended
to be reliable abundance indicators, and another Cu~\textsc{i} line at
2244.265~\AA\ is too weak.
The mean abundance derived from the two good Cu~\textsc{i} lines,
\logeps{Cu} = 0.47 $\pm$~0.09,
is considerably lower than the abundance derived 
by \citet{placco15cemps}
from the Cu~\textsc{i} line at 2824.370~\AA,
\logeps{Cu} = 2.20 $\pm$~0.20.
Reconsideration of the $\lambda$2824 line
in the lower-resolution E230M spectrum
analyzed previously suggests that most, if not all, of the absorption
at this wavelength may be due to an Fe~\textsc{i} line at 2824.394~\AA.~

Several Cu~\textsc{ii} lines are also detectable in our spectrum.
Four of them are reliable abundance indicators.
The line at 2112.100~\AA\ is 
blended with a Ni~\textsc{ii} line at 2112.107~\AA.~
This Ni~\textsc{ii} line cannot account for
all the absorption on the blue side of the line,
permitting us to derive a Cu abundance
from the Cu~\textsc{ii} line.
The Cu~\textsc{ii} lines at 2126.044 and 2189.630~\AA\ 
are well-resolved from nearby blending features,
as shown in Figures~\ref{specplot1} and \ref{specplot2}, respectively.
The Cu~\textsc{ii} line at 2247.002~\AA\ is blended with a stronger
Fe~\textsc{ii} line at 2246.916~\AA,
but the E230H spectrum mostly resolves this blend,
as shown in Figure~\ref{specplot3}.
Many other Cu~\textsc{ii} lines 
(at 2037.127, 2054.979, 2104.796, 2135.981, 
2148.983, 2179.410, 2192.268, 2210.267,
2218.108, 2228.867, 2242.618, 2276.258, and 2294.367~\AA)
are too blended or located in spectral regions with low S/N
to be useful as abundance indicators in \hdone.

The mean abundance derived from the two good Cu~\textsc{i} lines,
\logeps{Cu} = 0.47 $\pm$~0.09, 
is also lower than the mean abundance derived from the four good
Cu~\textsc{ii} lines,
\logeps{Cu} = 0.94 $\pm$~0.09.
Similar discrepancies have been noted previously
\citep{roederer12b,roederer14d,roederer16c,
andrievsky18,korotin18,
roederer18a,roederer22a}.
The magnitude of this offset, $\approx +0.5$~dex,
matches expectations for NLTE overionization of Cu in
metal-poor stars \citep{korotin18,shi18,caliskan25}.
The Cu abundance derived from Cu~\textsc{ii} lines 
should be reliable in LTE calculations, and we recommend this value
be used to indicate the Cu abundance in \hdone.

\subsection{\texorpdfstring{Gallium (Ga, $Z = 31$)}{Gallium (Ga, Z=31)}}
\label{gallium}

We detect the Ga~\textsc{ii} line at 2090.769~\AA\
in \hdone, as shown in Figure~\ref{specplot1}.
This line, which connects to the ground 
$3d^{10}4s^{2}$ $^{1}$S ground level,
has only been detected previously in the
\rpro-enhanced star \hdtwo\ by \citet{roederer22a}.
This line is blended with a pair of low-excitation 
Fe~\textsc{i} lines at 2090.855~\AA,
whose strengths can be fit to the observed line profile.

\subsection{\texorpdfstring{Germanium (Ge, $Z = 32$)}{Germanium (Ge, Z=32)}}
\label{germanium}

We derive Ge abundances from two Ge~\textsc{i} lines,
at 2094.258 and 2198.714~\AA.~
The $\lambda$2094 line is shown in Figure~\ref{specplot1}.
This line is located on the red wing of a strong Si~\textsc{i}
line at 2094.184~\AA, whose strength we fit to the observed spectrum.
The $\lambda$2198 line is weaker and detected on the red wing
of an Fe~\textsc{ii} line at 2198.665~\AA.~
NIST recommends a \loggf\ value of $-$0.47 for this line;
we reduce this value to $-$0.7, consistent with the D+
grade assigned by NIST, to better match the observed spectrum.
Absorption is detected at the wavelength of
the Ge~\textsc{i} line at 2065.215~\AA\ line,
but it falls in the damping wing of a strong Cr~\textsc{ii} resonance
line at 2065.501~\AA.~
We derive an upper limit from the $\lambda$2065 line
that is compatible with the abundances 
derived from the two detections.

\citet{placco15cemps} detected three Ge~\textsc{i} lines
in their E230M spectrum of \hdone.
We scale the \loggf\ values adopted there to the \citet{li99} scale
adopted here.
All five lines are included in
the weighted mean abundance reported in Table~\ref{abundtab}.

The FIP of Ge is 7.90~eV,
the same as for Fe,
and Ge is mostly ionized in the atmosphere of \hdone.
As for Fe, overionization of neutral Ge is likely.
NLTE calculations are not yet available for Ge lines in cool stars.
Until such calculations are available, we recommend that the 
abundances derived from the minority neutral species of Ge
be treated with caution, noting that they could be 
underestimated by a few tenths of a dex.

\subsection{\texorpdfstring{Arsenic (As, $Z = 33$)}{Arsenic (As, Z=33)}}
\label{arsenic}

We detect the As~\textsc{i} line at 2288.115~\AA\ in our
spectrum of \hdone.
This line could not be resolved from 
the stronger Cd~\textsc{i} line at 2288.023~\AA\ in 
previously examined STIS/E230M ($R = 30,000$) spectra.
Our STIS/E230H spectrum resolves it in \hdone,
as shown in Figure~\ref{specplot3}.
Other As~\textsc{i} lines that have been analyzed previously
in metal-poor stars are found at wavelengths 
shorter than those covered by our spectrum of \hdone.

\subsection{\texorpdfstring{Selenium (Se, $Z = 34$)}{Selenium (Se, Z=34)}}
\label{selenium}

The Se~\textsc{i} line at 2074.784~\AA\ is unblended 
in our spectrum of \hdone.
It lies in a region of relatively low S/N ($\approx$~15/1 pix$^{-1}$), 
as shown in Figure~\ref{specplot1}.
The Se~\textsc{i} line at 2062.779~\AA\ is too blended with
a much stronger Fe~\textsc{ii} line at 2062.788~\AA.~
The Se~\textsc{i} line at 2039.842~\AA\ is detected,
but it is found in a region with S/N $<$ 10/1 pix$^{-1}$
and is blended with a much stronger Cr~\textsc{ii} line at 2039.914~\AA.~
We derive an abundance from only the $\lambda$2074 line.

\subsection{\texorpdfstring{Molybdenum (Mo, $Z = 42$)}{Molybdenum (Mo, Z=42)}}
\label{molybdenum}

We checked several Mo~\textsc{ii} lines in 
our spectrum of \hdone, but none of them
are useful as abundance indicators.
Two lines, at 2038.452 and 2045.973~\AA, 
are located in 
regions of low S/N.~
Another line, at 2081.681~\AA, is strongly blended
with an unidentified line at $\approx$2081.63~\AA.~
We recommend the Mo abundance derived 
by \citet{placco15cemps} from one Mo~\textsc{ii} line at 2871.507~\AA.~

\subsection{\texorpdfstring{Ruthenium (Ru, $Z = 44$)}{Ruthenium (Ru, Z=44)}}
\label{ruthenium}

A weak absorption feature 
(depth $\approx$5\% of the continuum)
is present at the wavelength of 
the Ru~\textsc{ii} line at 2281.720~\AA.~
This feature is relatively unblended in \hdone,
as shown in Figure~\ref{specplot3}.
We derive an abundance from this line.
There are no other Ru~\textsc{ii} lines available
in our spectrum to validate this detection,
so caution is warranted.

\subsection{\texorpdfstring{Rhodium (Rh, $Z = 45$)}{Rhodium (Rh, Z=67)}}
\label{rhodium}

The FIP of Rh is low, 7.46~eV,
so most of the Rh is ionized in the atmosphere of \hdone.
Any lines of Rh~\textsc{i} are likely to be quite weak.
No absorption is detected at the wavelength of the Rh~\textsc{i} line
at 3434.89~\AA.~
We derive an upper limit on the Rh abundance from our MIKE spectrum
of \hdone.

\subsection{\texorpdfstring{Palladium (Pd, $Z = 46$)}{Palladium (Pd, Z=46)}}
\label{palladium}

The FIP of Pd, 8.34~eV, is moderately higher than that of Rh,
so some neutral Pd may be present.
A weak Pd~\textsc{i} absorption line is detectable at 3404.58~\AA\ in our 
MIKE spectrum of \hdone.
This line is located between stronger lines of Fe~\textsc{i} at
3404.30~\AA\ and Zr~\textsc{ii} at 3404.83~\AA.~
We derive a Pd abundance from this line,
and we conclude that a large observational uncertainty, 
$\approx$~0.25~dex,
is warranted.

\subsection{\texorpdfstring{Silver (Ag, $Z = 47$)}{Silver (Ag, Z=47)}}
\label{silver}

The FIP of Ag, like Rh, is low, 7.58~eV,
and most Ag is ionized in the atmosphere of \hdone.
No Ag~\textsc{i} absorption is detectable at 3382.89~\AA\ in our
MIKE spectrum of \hdone.
Weak absorption, with continuum depth $<$~10\%, 
is likely due to NH at this wavelength.
We derive an upper limit on the Ag abundance assuming 
this absorption is caused by Ag~\textsc{i}.

\subsection{\texorpdfstring{Cadmium (Cd, $Z = 48$)}{Cadmium (Cd, Z=48)}}
\label{cadmium}

The Cd~\textsc{i} line at 2288.023~\AA\
is strong in \hdone, and it dominates the 
absorption at this wavelength,
as shown in Figure~\ref{specplot3}.
The abundance derived from this line,
\logeps{Cd} = $-$0.67 $\pm$~0.10,
agrees with that derived 
by \citet{placco15cemps} from the same line
in the STIS/E230M spectrum,
\logeps{Cd} = $-$0.70 $\pm$~0.20.
The Cd~\textsc{ii} line at 2144.390~\AA\
is part of a large absorption complex that includes
Fe~\textsc{i} lines at 2144.350 and 2144.449~\AA.~
Neither of these Fe~\textsc{i} lines has an experimental \loggf\ value,
and we adjust their strengths to match the observed line profile.
We report a Cd abundance from the $\lambda$2144 line with caution
given these blends.
The Cd~\textsc{ii} line at 2265.020~\AA\ is blended with
a much stronger Fe~\textsc{i} line at 2265.054~\AA,
and we do not consider this line further.
The two acceptable Cd lines yield abundances that are in agreement.

The FIP of Cd is 8.99~eV, and 
a substantial fraction of Cd is found in each of the
neutral and ionized states.
NLTE calculations are not yet available for Cd in cool stars.
\citet{shah24} found that the \logeps{Cd/Zr} ratio 
in eight \rpro-enhanced stars varies by $\approx$1~dex,
and it correlates closely with \teff\ and \logg.
This ratio is not expected to exhibit large variations 
in such stars.
\citeauthor{shah24}\ concluded that the 
Cd~\textsc{i} line at 2288.023~\AA,
which most of these results are based on,
may not be formed in LTE.~
Our abundance derivations from both neutral and ionized Cd
yield abundances that agree to within $\approx$0.3~dex.
We urge caution when interpreting the Cd abundance in
\hdone\ and amplify the call by \citeauthor{shah24} 
for new NLTE calculations of Cd abundances in metal-poor stars.

\subsection{\texorpdfstring{Indium (In, $Z = 49$)}{Indium (In, Z=49)}}
\label{indium}

The FIP of In is 5.79~eV, so virtually all In is ionized 
in the atmosphere of \hdone.
Only one line of In~\textsc{ii} is potentially detectable 
in the UV spectra of \hdone, 2306.064~\AA.~
Nevertheless, any absorption at this wavelength is minimal.
We derive an upper limit on the In abundance using this line
in our STIS/E230M spectrum of \hdone.

\subsection{\texorpdfstring{Tin (Sn, $Z = 50$)}{Tin (Sn, Z=50)}}
\label{tin}

The FIP of Sn is 7.34~eV, and most Sn is ionized in the
atmosphere of \hdone.
We detect the Sn~\textsc{ii} line at 2151.514~\AA.~
This line is unblended in \hdone,
as shown in Figure~\ref{specplot2}.
No Sn~\textsc{i} lines are detected.
Upper limits derived from the Sn~\textsc{i} lines at
2199.346 and 2286.681~\AA,
\logeps{Sn} $<$ 0.8, are
consistent with the abundance derived from the 
Sn~\textsc{ii} line detected at 2151~\AA.~

\subsection{\texorpdfstring{Antimony (Sb, $Z = 51$)}{Antimony (Sb, Z=51)}}
\label{antimony}

We search for two Sb~\textsc{i} lines,
at 2068.344 and 2175.818~\AA.~
The line at 2175~\AA\ is in an otherwise
clean region of the continuum, yet we detect no absorption from
Sb~\textsc{i}, as shown in Figure~\ref{specplot2}.
We derive an upper limit of \logeps{Sb} $< -$1.1 from this line.
Weak absorption is detected at the $\lambda$2068 line.
This feature lies on the blue wing of a much stronger 
Cr~\textsc{ii} line at 2068.395~\AA, which we 
adjust to fit the observed line profile.
The abundance inferred from the $\lambda$2068 line,
\logeps{Sb} $\approx -$0.9,
is higher than the upper limit inferred from the $\lambda$2175 line.
We conclude that the absorption at 2068~\AA\ is not due to 
Sb~\textsc{i}, and we instead report an upper limit.

\subsection{\texorpdfstring{Tellurium (Te, $Z = 52$)}{Tellurium (Te, Z=52)}}
\label{tellurium}

We detect the Te~\textsc{i} line at 2259.034~\AA\
in our spectrum of \hdone, as shown in Figure~\ref{specplot3}.
The line profile fit is a bit uneven, but this line is 
otherwise unblended.
The Te~\textsc{ii} line at 2142.822~\AA\ is blended with a number of
strong lines of other species, including Ti~\textsc{ii},
V~\textsc{ii}, Fe~\textsc{i}, and Fe~\textsc{ii}.
These other species can fully account for the absorption
at this wavelength in \hdone.
We rechecked the Te~\textsc{i} line at 2385.792~\AA\
in the STIS/E230M spectrum of \hdone,
and we infer an upper limit of \logeps{Te} $< +$0.3.
This upper limit is compatible with the abundance derived from the
$\lambda$2259 line, \logeps{Te} $= -$0.68 $\pm$~0.25.

\subsection{\texorpdfstring{Holmium (Ho, $Z = 67$)}{Holmium (Ho, Z=67)}}
\label{holmium}

We detect absorption, with $\approx$~10\% continuum depth,
at the wavelength of the 
Ho~\textsc{ii} line at 3484.83~\AA\ in our
MIKE spectrum of \hdone.
This line is located on the blue wing of a 
stronger Fe~\textsc{i} line at 3484.98~\AA,
and our synthetic spectrum provides an imperfect fit to the
observed line profile.
We are confident that Ho~\textsc{ii} absorption is present,
but we assign a relatively large uncertainty to the Ho abundance.

\subsection{\texorpdfstring{Thulium (Tm, $Z = 69$)}{Thulium (Tm, Z=69)}}
\label{thulium}

The Tm~\textsc{ii} line at 3462.20~\AA\ is detectable in our
MIKE spectrum of \hdone.
It is mildly blended with an Fe~\textsc{i} line at 3462.35~\AA\
of roughly equal strength.
These two lines are sufficiently resolved to permit
a reliable Tm abundance determination.

\subsection{\texorpdfstring{Ytterbium (Yb, $Z = 70$)}{Ytterbium (Yb, Z=70)}}
\label{ytterbium}

Two Yb~\textsc{ii} lines yield abundances in \hdone.
The line at 2116.675~\AA\ is unblended, and 
the line at 2126.741~\AA\ is mildly blended with an Fe~\textsc{i} line.
Both lines are shown in Figure~\ref{specplot2}.
The abundances derived from these two lines differ by 0.41~dex,
\logeps{Yb} = $-0.59 \pm 0.14$ and $-1.00 \pm 0.25$, respectively.
\citet{roederer14c} reported the Yb abundance in \hdone\
from an optical Yb~\textsc{ii} line at 3694~\AA,
and its abundance agrees, 
\logeps{Yb} = $-0.68 \pm 0.20$.
All three Yb~\textsc{ii} lines arise from the 
$4f^{16}6s$ $^{2}$S ground state.
These three lines are used in computing the weighted mean Yb abundance
listed in Table~\ref{abundtab}.

\subsection{\texorpdfstring{Lutetium (Lu, $Z = 71$)}{Lutetium (Lu, Z=71)}}
\label{lutetium}

The Lu~\textsc{ii} line at 2195.556~\AA\ is detected in our spectrum
of \hdone.
This line is unblended, as shown in Figure~\ref{specplot2}.
The observed line is well-fit by our synthetic spectrum, but
we caution that several weak unidentified absorption features
are found nearby.
The Lu abundance reported in Table~\ref{abundtab} 
includes the abundance derived by \citet{placco15cemps}
from the Lu~\textsc{ii} $\lambda$2615 line 
in the STIS/E230M spectrum of \hdone.
The abundances derived from the two lines are in good agreement:\
\logeps{Lu} = $-$1.85 $\pm$~0.15 from the line at 2195~\AA, and
\logeps{Lu} = $-$1.90 $\pm$~0.20 from the line at 2615.412~\AA.~

\subsection{\texorpdfstring{Hafnium (Hf, $Z = 72$)}{Hafnium (Hf, Z=72)}}
\label{hafnium}

\citet{denhartog21hf} derived Hf abundances from 12 Hf~\textsc{ii}
lines in UV and optical spectra of \hdone.
Previously, only one Hf~\textsc{ii} line had been 
available for the \citet{placco15cemps} study.
We adopt the Hf abundance from \citeauthor{denhartog21hf} 

\subsection{\texorpdfstring{Tungsten (W, $Z = 74$)}{Tungsten (W, Z=74)}}
\label{tungsten}

Five W~\textsc{ii} lines are detected in our 
spectrum of \hdone.
The W~\textsc{ii} lines at 2088.204 and 2118.874~\AA\ are unblended, 
as shown in Figure~\ref{specplot1}.
The W~\textsc{ii} line at 2094.751~\AA\ is located between
an Fe~\textsc{ii} line at 2094.634~\AA\ and 
an Al~\textsc{i} line at 2094.847~\AA.~
Both of these blending features can be fit,
as also shown in Figure~\ref{specplot1}.
The W~\textsc{ii} line at 2194.528~\AA\ is blended only by
a weak Co~\textsc{ii} line at 2194.446~\AA\
and an Os~\textsc{ii} line at 2194.403~\AA\ (see Section~\ref{osmium}),
as shown in Figure~\ref{specplot2}.
Finally, the W~\textsc{ii} line at 2248.758~\AA\ is detected 
between a Co~\textsc{ii} line at 2248.667~\AA\ and an
Fe~\textsc{i} line at 2248.860~\AA,
as shown in Figure~\ref{specplot3}.

Absorption is also detected at 2204.489~\AA, which matches the 
wavelength of a W~\textsc{ii} line with E.P.\ of 0.76~eV,
as predicted by the \citet{kurucz11} lists and the NIST ASD.~
There are no other species with plausible transitions predicted
at this wavelength.
No laboratory study has measured the lifetime of the upper level
of this transition, 51495.054~cm$^{-1}$.
We assume that the absorption at this wavelength is entirely W~\textsc{ii},
and we use the W abundance derived from the five other W~\textsc{ii}
lines to estimate the \loggf\ value of this line,
$-$0.10, with an uncertainty of about 0.15~dex.
We cannot, of course, derive an abundance from this line in \hdone,
but it may be of use in other stars.

\subsection{\texorpdfstring{Rhenium (Re, $Z = 75$)}{Rhenium (Re, Z=75)}}
\label{rhenium}

We tentatively detect two Re~\textsc{ii} lines,
at 2214.277 and 2275.255~\AA, in our spectrum of \hdone.
The $\lambda$2214 line, shown in Figure~\ref{specplot2},
is blended with Fe~\textsc{i} lines at 2214.257 and 2214.295~\AA\
and a Cr~\textsc{i} line at 2214.330~\AA,
all of which have comparable strength to the Re~\textsc{ii} line.
None of these blending features have laboratory \loggf\ values,
so we empirically constrain them
using the observed line profile
(\loggf\ $\approx$ $-$2.8, $-$4.8, and $-$2.5, respectively).
There is no way to fully account for the absorption near this wavelength
using only these blending lines without overpredicting the 
absorption near the center and underpredicting the absorption near the wings.
We thus simultaneously fit these blending features
and the Re abundance.
The wide HFS of the Re~\textsc{ii} line, $\approx$~0.15~\AA,
extends on both sides of the blending lines.

The Re~\textsc{ii} line at 2275.255~\AA\ is
shown in Figure~\ref{specplot3}.
There is a weak Cr~\textsc{i} line at 2275.313~\AA\ in the 
\citet{kurucz11} line list that could potentially 
contribute some fraction of the absorption.
Its \loggf\ value is not known experimentally.
This Cr~\textsc{i} line alone cannot account for the wide
absorption observed at this wavelength, even
if we artificially increase its \loggf\ value.
The Re abundance derived from this line,
\logeps{Re} = $-$1.56 $\pm$~0.25, 
agrees with that derived from the $\lambda$2214 line,
\logeps{Re} = $-$1.69 $\pm$~0.25.

\subsection{\texorpdfstring{Osmium (Os, $Z = 76$)}{Osmium (Os, Z=76)}}
\label{osmium}

Two Os~\textsc{ii} lines are detected at 2194.403, and 2282.278~\AA,
as shown in Figures~\ref{specplot2} and \ref{specplot3}, respectively.
The $\lambda$2282 line was also examined by \citet{placco15cemps},
and our abundance supersedes the upper limit derived previously
from the lower-S/N STIS/E230M spectrum.
\citet{roederer22a} recommended against using the
Os~\textsc{ii} line at 2067.230~\AA\
until further study of its blending feature(s)
becomes available, so we do not use this line 
to derive an Os abundance.
The Os~\textsc{ii} line at 2164.838~\AA\ is not detected,
and we report an upper limit from this line.
Weak absorption is detected at the wavelength of 
the Os~\textsc{ii} line at 2227.980~\AA.~
There are several unidentified absorption lines in this region
that complicate the continuum placement, so we report only an 
upper limit from this line.
Both of these upper limits are compatible with the abundances
derived from the two Os~\textsc{ii} lines that are detected.

\subsection{\texorpdfstring{Iridium (Ir, $Z = 77$)}{Iridium (Ir, Z=77)}}
\label{iridium}

There are no clear detections of Ir~\textsc{ii} lines
at 2221.068, 2245.750, or 2255.406~\AA\ in our spectrum of \hdone.
Weak absorption is detected at 2221.068~\AA.~
If we assume that this absorption is due to Ir~\textsc{ii}, the 
derived abundance is about 0.5~dex higher than
the upper limit derived from the $\lambda$2245 line,
as shown in Figure~\ref{specplot3}.
We thus consider both lines to be non-detections of Ir~\textsc{ii}.
As discussed in \citet{roederer14d}, absorption detected at 
$\approx$2126.82~\AA\
is not likely due to Ir~\textsc{ii}.

\subsection{\texorpdfstring{Platinum (Pt, $Z = 78$)}{Platinum (Pt, Z=78)}}
\label{platinum}

We detect two Pt~\textsc{i} lines, at 2144.212 and 2274.381~\AA,
as shown in Figures~\ref{specplot2} and \ref{specplot3}, respectively.
The former line has a weak blend with an unidentified line
at $\approx$2144.255~\AA,
which can be accounted for in the synthesis.
An upper limit inferred from the non-detection of Pt~\textsc{i} 
at 2067.509~\AA\ is compatible with the two detections.
Absorption at 2165.211~\AA\ matches the wavelength of
another Pt~\textsc{i} line, but the abundance inferred from this
line is greater than that derived from the upper limit from the
Pt~\textsc{i} line at 2067~\AA, so we conclude that
the absorption here is also not due to Pt~\textsc{i}.
The S/N is too low to infer a platinum abundance from the
Pt~\textsc{i} line at 2049.391~\AA, 
and other Pt~\textsc{i} lines at 2103.344 and 2274.841~\AA\
are too weak or blended to derive meaningful upper limits.
The mean Pt abundance listed in Table~\ref{abundtab}
includes the abundance derived 
from the Pt~\textsc{i} $\lambda$2659 line 
from \citet{placco15cemps}.

\subsection{\texorpdfstring{Lead (Pb, $Z = 82$)}{Lead (Pb, Z=82)}}
\label{lead}

\citet{roederer20} derived the Pb abundance
from the Pb~\textsc{ii} line at 2203.534~\AA\
in our spectrum of \hdone.
The Pb~\textsc{ii} line arises from the low-excitation
$^{2}P^{\rm o}_{3/2}$ level (E.P.\ = 1.75~eV).
It is saturated in our spectrum of \hdone, 
even after accounting for the de-saturating effects of HFS and IS.~
Nevertheless, the line profile is well-fit by our synthetic spectra.
\citet{peterson21} postulated that a weak
Fe~\textsc{i} line at 2203.526~\AA\
could account for absorption observed
at this wavelength in spectra of 
other metal-poor stars
without strong Pb enhancement.
Including that proposed Fe~\textsc{i} line
in our synthesis would decrease the derived Pb abundance
by $< 0.1$~dex.

The abundance derived from the $\lambda$2203 line,
\logeps{Pb} = 1.91 $\pm$~0.30,
is considerably higher than that derived from
Pb~\textsc{i} lines at 2833 and 4057~\AA,
\logeps{Pb} = 1.35 $\pm$~0.20 and 1.50 $\pm$~0.20, respectively
\citep{placco15cemps}.
This $\approx$~0.5~dex difference is 
potentially explained by NLTE,
because this difference matches
the approximate NLTE correction to the LTE abundance derived from the
Pb~\textsc{i} line at 4057~\AA\ \citep{mashonkina12}.
The FIP of Pb is 7.42~eV, so Pb is almost completely ionized
in the atmosphere of \hdone.
\citeauthor{mashonkina12}\
showed that the ground state of ionized Pb
is reliably predicted by LTE calculations,
and no calculations are yet available for
excited levels of ionized Pb.
We recommend adopting the Pb abundance derived from the 
$\lambda$2203 Pb~\textsc{ii} line in \hdone.

\subsection{\texorpdfstring{Bismuth (Bi, $Z = 83$)}{Bismuth (Bi, Z=83)}}
\label{bismuth}

We checked two Bi~\textsc{i} lines in our spectrum of \hdone, 
and we detect neither one of them.
As shown in Figure~\ref{specplot3},
the Bi~\textsc{i} line at 2230.609~\AA\ is located in a region
between several other weak lines, including a
Cr~\textsc{ii} line at 2230.571~\AA\ and 
Fe~\textsc{ii} lines at 2230.628 and 2230.679~\AA.~
None of these blending features has a laboratory \loggf\ value.
The observed line profile can be reasonably well-fit
with no Bi in the synthetic spectrum
when we empirically adjust the \loggf\ values of these blending features
(\loggf\ = $-$2.8, $-$3.0, and $-$2.45, respectively).
We conclude that Bi~\textsc{i} is not detected at this wavelength.
The synthesis shown in Figure~\ref{specplot3}
adopts reduced values for the \loggf\ values of these blending features
and the maximum amount of Bi~\textsc{i} that could plausibly
account for the observed line profile.

A second Bi~\textsc{i} line at 2276.544~\AA\ is much weaker and
blended with a stronger, unidentified feature at $\approx$2276.53~\AA.~
We are unable to derive an upper limit from this line.

The FIP of Bi is 7.29~eV, and 
nearly all Bi atoms are ionized in the atmosphere of \hdone.
NLTE calculations are not yet available.
Until such calculations are available, we recommend that the 
upper limit derived from the minority neutral species of Bi
be treated with caution, noting that it could be 
underestimated by a few tenths of a dex.

\bibliographystyle{aasjournalv7}
\bibliography{ms.bbl}

\end{document}

%% file: tab1-stub.tex
\begin{deluxetable}{ccccccc}
\tablecaption{Line Atomic Data, References, and Derived Abundances
\label{linetab}}
\tabletypesize{\scriptsize}
\tablehead{
\colhead{Species} &
\colhead{$\lambda$} &
\colhead{E.P.} &
\colhead{\loggf} &
\colhead{Ref.} &
\colhead{$\log\varepsilon$} &
\colhead{Unc.} \\
\colhead{} &
\colhead{(\AA)} &
\colhead{(eV)} &
\colhead{} &
\colhead{} &
\colhead{} &
\colhead{(dex)} 
}
\startdata
Fe~\textsc{i}  & 2132.017 &  0.00 & $-$1.67 & 1  &     5.30 &   0.21   \\
Fe~\textsc{i}  & 2141.718 &  0.05 & $-$2.25 & 1  &     5.33 &   0.25   \\
Fe~\textsc{i}  & 2176.840 &  0.12 & $-$1.66 & 2  &     5.42 &   0.22   \\
\enddata
\tablereferences{%
 1 = \citet{belmonte17};
 2 = NIST \citep{kramida21};
 3 = \citet{denhartog19};
 4 = \citet{morton03};
 5 = \citet{roederer12b};
 6 = \citet{roederer22a};
 7 = \citet{li99};
 8 = \citet{holmgren75}, using the line component pattern presented in 
\citet{roederer22a};
 9 = \citet{morton00};
10 = \citet{johansson94};
11 = \citet{duquette85};
12 = \citet{hansen12} for both the \loggf\ value and line component pattern;
13 = \citet{xu04};
14 = \citet{curtis00in}, using the line component pattern presented in 
\citet{roederer22a};
15 = \citet{oliver10};
16 = \citet{hartman10}, using the line component patterns presented in 
\citet{roederer22a};
17 = \citet{roederer12a};
18 = \citet{lawler04} for both the \loggf\ value and line component pattern;
19 = \citet{wickliffe97tm};
20 = \citet{quinet99}, using the line component pattern presented in 
\citet{denhartog20};
21 = \citet{nilsson08}, using the line component patterns presented in 
\citet{roederer22a};
22 = \citet{kling00}, using the line component patterns presented in 
\citet{roederer22a};
23 = \citet{palmeri05re}, using the line component pattern presented in 
\citet{roederer22a};
24 = Based on \citet{palmeri05re} and \citet{wahlgren97}, using the 
\loggf\ value and line component pattern presented in \citet{roederer22a};
25 = \citet{quinet06};
26 = \citet{ivarsson04};
27 = \citet{xu07};
28 = \citet{denhartog05};
29 = \citet{roederer14d};
30 = \citet{denhartog05}, using the line component pattern presented in
\citet{roederer22a};
31 = \citet{quinet07}, using the line component pattern presented in 
\citet{roederer20}.
}
\tablecomments{%
A complete machine-readable version of Table~\ref{linetab} is
available online.
A short version is shown here to illustrate its form and content.}
\end{deluxetable}

%% file: tab2.tex
\begin{deluxetable}{ccccccc}
\tablecaption{Mean Abundances in \hdone\ Derived in This Study
\label{abundtab}}
\tabletypesize{\scriptsize}
\tablehead{
\colhead{Species} &
\colhead{$\log\varepsilon_{\odot}$\tablenotemark{a}} &
\colhead{$\log\varepsilon$} &
\colhead{[X/Fe]} &
\colhead{Unc.\tablenotemark{b}} &
\colhead{N$_{\rm lines}$} &
\colhead{Notes} \\
\colhead{} &
\colhead{} &
\colhead{} &
\colhead{} &
\colhead{(dex)} &
\colhead{} &
\colhead{}
}
\startdata
B~\textsc{i}   & 2.81 & $<-$0.60 & $<-$1.00 & \nodata & 2 & (\tablenotemark{c}) \\ 
P~\textsc{i}   & 5.45 &     2.80 &  $-$0.24 &    0.23 & 2 &  \\ 
Cu~\textsc{i}  & 4.27 &     0.47 &  $-$1.39 &    0.09 & 2 &  \\
Cu~\textsc{ii} & 4.27 &     0.94 &  $-$0.92 &    0.09 & 4 &  \\
Ga~\textsc{ii} & 3.10 &     0.30 &  $-$0.39 &    0.30 & 1 &  \\
Ge~\textsc{i}  & 3.59 &     0.36 &  $-$0.82 &    0.10 & 5 & (\tablenotemark{d}) \\ 
As~\textsc{i}  & 2.32 &  $-$0.37 &  $-$0.28 &    0.18 & 1 &  \\ 
Se~\textsc{i}  & 3.36 &     1.30 &  $+$0.35 &    0.25 & 1 &  \\ 
Ru~\textsc{ii} & 1.78 &  $-$0.21 &  $+$0.42 &    0.20 & 1 &  \\
Rh~\textsc{i}  & 1.10 & $<-$0.30 & $<+$1.01 & \nodata & 1 &  \\
Pd~\textsc{i}  & 1.67 &  $-$0.40 &  $+$0.34 &    0.25 & 1 &  \\
Ag~\textsc{i}  & 1.22 & $<-$0.65 & $<+$0.54 & \nodata & 1 &  \\
Cd~\textsc{i}  & 1.73 &  $-$0.67 &  $+$0.01 &    0.13 & 1 &  \\
Cd~\textsc{ii} & 1.73 &  $-$0.97 &  $-$0.29 &    0.25 & 1 &  \\
In~\textsc{ii} & 0.78 & $<-$0.60 & $<+$1.03 & \nodata & 1 &  \\
Sn~\textsc{ii} & 2.09 &  $-$0.04 &  $+$0.28 &    0.22 & 1 &  \\
Sb~\textsc{i}  & 1.03 & $<-$1.10 & $<+$0.28 & \nodata & 2 &  \\ 
Te~\textsc{i}  & 2.20 &  $-$0.68 &  $-$0.47 &    0.25 & 1 &  \\ 
Ho~\textsc{ii} & 0.49 &  $-$1.70 &  $+$0.22 &    0.20 & 1 &  \\
Tm~\textsc{ii} & 0.14 &  $-$1.64 &  $+$0.63 &    0.20 & 1 &  \\
Yb~\textsc{ii} & 0.94 &  $-$0.69 &  $+$0.78 &    0.11 & 3 &  \\ 
Lu~\textsc{ii} & 0.11 &  $-$1.87 &  $+$0.43 &    0.12 & 2 &  \\ 
Hf~\textsc{ii} & 0.73 &  $-$0.72 &  $+$0.96 &    0.09 &12 & (\tablenotemark{e}) \\
W~\textsc{ii}  & 0.67 &  $-$0.86 &  $+$0.88 &    0.08 & 5 &  \\ 
Re~\textsc{ii} & 0.28 &  $-$1.63 &  $+$0.50 &    0.18 & 2 &  \\
Os~\textsc{ii} & 1.37 &  $-$0.88 &  $+$0.16 &    0.11 & 2 &  \\
Ir~\textsc{ii} & 1.36 & $<$ 0.20 & $<+$1.25 & \nodata & 2 &  \\ 
Pt~\textsc{i}  & 1.64 &  $-$0.69 &  $+$0.08 &    0.15 & 3 &  \\
Pb~\textsc{ii} & 2.06 &     1.91 &  $+$2.26 &    0.23 & 1 & (\tablenotemark{f}) \\ 
Bi~\textsc{i}  & 0.67 & $<-$0.50 & $<+$1.24 & \nodata & 1 & (\tablenotemark{d}) \\ 
\enddata
\tablecomments{%
Abundances of elements not reported in Table~\ref{abundtab}
may be found in the references cited in Section~\ref{results}.
}
\tablenotetext{a}{%
\citet{lodders09}
}
\tablenotetext{b}{%
The uncertainty listed here is the measurement uncertainty,
which includes the individual line fitting uncertainties and
the \loggf\ uncertainties.
Systematic uncertainties that account for uncertainties 
in the model atmosphere parameters are $\sim$~0.2~dex
(see Table~6 of \citealt{placco15cemps}).
}
\tablenotetext{c}{%
LTE abundance upper limit given.
NLTE corrections may increase these upper limits by $\approx+0.9$~dex;
See Appendix~\ref{boron} and \citet{kiselman96}.}
\tablenotetext{d}{%
NLTE calculations are not available but could be expected
to yield abundances a few tenths of a dex higher than the LTE abundance 
given here.}
\tablenotetext{e}{%
From \citet{denhartog21hf}}
\tablenotetext{f}{%
From \citet{roederer20}}
\end{deluxetable}

%% file: tab3.tex
\begin{deluxetable}{ccrrrcc}
\tablecaption{Recommended Heavy-element Abundances in \hdone
\label{finaltab}}
\tabletypesize{\small}
\tabletypesize{\scriptsize}
\tablehead{
\colhead{Species} &
\colhead{$Z$} &
\colhead{$\log\varepsilon_{\odot}$} &
\colhead{$\log\varepsilon$} &
\colhead{[X/Fe]} &
\colhead{Unc.} &
\colhead{Ref.} \\
\colhead{} &
\colhead{} &
\colhead{} &
\colhead{} &
\colhead{} &
\colhead{(dex)} &
\colhead{}
}
\startdata
Ga &   31 &     3.10 &         0.30  &  $-$0.39  &  0.30    & 1 \\
Ge &   32 &     3.59 &         0.36  &  $-$0.82  &  0.20    & 1 \\
As &   33 &     2.32 &      $-$0.37  &  $-$0.28  &  0.20    & 1 \\
Se &   34 &     3.36 &         1.30  &  $+$0.35  &  0.25    & 1 \\
Rb &   37 &     2.38 &     $<$ 2.05  & $<+$2.08  &  \nodata & 2 \\
Sr &   38 &     2.90 &         1.25  &  $+$0.76  &  0.20    & 3 \\
Y  &   39 &     2.20 &         0.38  &  $+$0.59  &  0.22    & 3 \\
Zr &   40 &     2.57 &         0.89  &  $+$0.73  &  0.20    & 3 \\
Nb &   41 &     1.42 &      $-$0.70  &  $+$0.29  &  0.20    & 3 \\
Mo &   42 &     1.94 &      $-$0.30  &  $+$0.17  &  0.20    & 3 \\
Ru &   44 &     1.78 &      $-$0.21  &  $+$0.42  &  0.20    & 1 \\
Rh &   45 &     1.10 &     $<-$0.30  & $<+$1.01  &  \nodata & 1 \\
Pd &   46 &     1.67 &      $-$0.40  &  $+$0.34  &  0.25    & 1 \\
Ag &   47 &     1.22 &     $<-$0.65  & $<+$0.54  &  \nodata & 1 \\
Cd &   48 &     1.73 &      $-$0.73  &  $-$0.05  &  0.20    & 1 \\
In &   49 &     0.78 &     $<-$0.60  & $<+$1.03  &  \nodata & 1 \\
Sn &   50 &     2.09 &      $-$0.04  &  $+$0.28  &  0.22    & 1 \\
Sb &   51 &     1.03 &     $<-$1.10  & $<+$0.28  &  \nodata & 1 \\
Te &   52 &     2.20 &      $-$0.68  &  $-$0.47  &  0.25    & 1 \\
Ba &   56 &     2.18 &         1.00  &  $+$1.23  &  0.20    & 3 \\
La &   57 &     1.19 &      $-$0.37  &  $+$0.85  &  0.20    & 2 \\
Ce &   58 &     1.60 &         0.25  &  $+$1.06  &  0.20    & 2 \\
Pr &   59 &     0.77 &      $-$0.75  &  $+$0.89  &  0.20    & 2 \\
Nd &   60 &     1.47 &         0.07  &  $+$1.01  &  0.20    & 2 \\
Sm &   62 &     0.96 &      $-$0.73  &  $+$0.72  &  0.20    & 2 \\
Eu &   63 &     0.53 &      $-$1.75  &  $+$0.13  &  0.20    & 2 \\
Gd &   64 &     1.09 &      $-$0.59  &  $+$0.73  &  0.20    & 2 \\
Tb &   65 &     0.34 &     $<-$1.00  & $<+$1.07  &  \nodata & 2 \\
Dy &   66 &     1.14 &      $-$0.77  &  $+$0.50  &  0.20    & 2 \\
Ho &   67 &     0.49 &      $-$1.70  &  $+$0.22  &  0.20    & 1 \\
Er &   68 &     0.95 &      $-$0.80  &  $+$0.66  &  0.20    & 2 \\
Tm &   69 &     0.14 &      $-$1.64  &  $+$0.63  &  0.20    & 1 \\
Yb &   70 &     0.94 &      $-$0.69  &  $+$0.78  &  0.20    & 1 \\
Lu &   71 &     0.11 &      $-$1.87  &  $+$0.43  &  0.20    & 1 \\
Hf &   72 &     0.73 &      $-$0.72  &  $+$0.96  &  0.20    & 4 \\
W  &   74 &     0.67 &      $-$0.86  &  $+$0.88  &  0.20    & 1 \\
Re &   75 &     0.28 &      $-$1.63  &  $+$0.50  &  0.20    & 1 \\
Os &   76 &     1.37 &      $-$0.88  &  $+$0.16  &  0.20    & 1 \\
Ir &   77 &     1.36 &     $<$ 0.20  & $<+$1.25  &  \nodata & 1 \\
Pt &   78 &     1.64 &      $-$0.69  &  $+$0.08  &  0.20    & 1 \\
Au &   79 &     0.82 &      $-$1.60  &  $-$0.01  &  0.20    & 3 \\
Pb &   82 &     2.06 &         1.91  &  $+$2.26  &  0.30    & 5 \\
Bi &   83 &     0.67 &     $<-$0.50  & $<+$1.24  &  \nodata & 1 \\
Th &   90 &     0.02 &     $<-$1.11  & $<+$1.28  &  \nodata & 2 \\
\enddata
\tablecomments{%
\citet{placco15cemps}\ did not list 
uncertainties for abundances derived from individual lines,
so we adopt a typical uncertainty of 0.20~dex
for each line in their study.
\citet{roederer14c} used slightly different model atmosphere parameters,
so we scale those abundances to the scale adopted here,
which was established by \citeauthor{placco15cemps} 
To do so, we scale the
$\log\varepsilon$ abundances of \citeauthor{roederer14c}\
using the [X/Fe] ratios.
}
\tablereferences{%
1 = This study;
2 = \citet{roederer14c};
3 = \citet{placco15cemps};
4 = \citet{denhartog21hf};
5 = \citet{roederer20}.
}
\end{deluxetable}